A SECURITY PRICE VOLATILE TRADING CONDITIONING MODEL

Leilei Shi*[1], Yiwen Wang[2], Ding Chen[3], Liyan Han[2], Yan Piao, and Chengling Gou[4]

[1]Complex System Research Group, Department of Modern Physics
University of Science and Technology of China
[2]Department of Finance, Beijing University of Aeronautics and Astronautics
[3]Harvest Fund Management Co. Ltd.
[4]Department of Physics, Beijing University of Aeronautics and Astronautics

Abstract

We develop a theoretical trading conditioning model subject to price volatility and return information in terms of market psychological behavior, based on analytical transaction volume-price probability wave distributions in which we use transaction volume probability to describe price volatility uncertainty and intensity. Applying the model to high frequent data test in China stock market, we have main findings as follows: 1) there is, in general, significant positive correlation between the rate of mean return and that of change in trading conditioning intensity; 2) it lacks significance in spite of positive correlation in two time intervals right before and just after bubble crashes; and 3) it shows, particularly, significant negative correlation in a time interval when SSE Composite Index is rising during bull market. Our model and findings can test both disposition effect and herd behavior simultaneously, and explain excessive trading (volume) and other anomalies in stock market.

*Corresponding author. E-mail address: Shileilei8@yahoo.com.cn or leilei.shi@hotmail.com, working in an international securities company in China.
The author appreciates valuable discussion with Huaiyu Wang, Yaping Wang, Chunsheng Zhou, Jing Liu, and Famin Liu etc.

## 1. INTRODUCTION

It has been long that literature in finance focuses much more attention on price and return and less on trading volume, even completely ignoring it. In the past 10 years, however, there is a changing trend that academics have increasing minds on the information contained in trading volume. In neoclassical finance framework, Lo and Wang (2006) explored the link between the dynamic properties of volume, together with price, and the economic fundamentals, and underlined the general point that trading volume and price should be an integral part of any analysis of asset market. In newly emerging behavioral finance, we have associated trading volume with investors' emotion, belief, and preference. Behavioral finance is the study of the influence of psychology on the behavior of financial practitioners and the subsequent effect on markets (Sewell, 2008). It helps explain why and how markets might be inefficient. Lee and Swaminathan (2000) showed that past trading volume provides an important link between "momentum" and "value" strategies and these findings help to reconcile intermediate-horizon "underreaction" and long-horizon "overreaction" effects. Benos (1998) and Odean (1998) hypothesized that overconfidence produces excessive trading in stock market. Odean (1999) explained why those actively trade in financial markets to be more overconfident than the general population by three reasons: selection bias, survivorship bias, and unrealistic belief, and tested overconfident trading hypothesis by investigating whether the trading profits of discount brokerage customers are sufficient to cover their trading costs. Barber and Odean (2000) documented further evidence that active trading results poor performance and is hazardous to wealth. Such irrational behavior can be explained only by overconfidence. Graham et al. (2009) found that investors who feel competent trade more, and thus explained that overconfidence leads to higher trading frequency. Tested the trading volume predictions of formal overconfidence models, Statman et al. (2006) found that share turnover is positively related to lagged returns in both market-wide and individual security for many months. They are interpreted as the evidence of investor overconfidence and the disposition effect. Barber et al. (2009) demonstrated that psychological biases that lead investors to systematically buy stocks with strong recent performance, to refrain from selling stocks with a loss, and to be net buyers of stocks with unusually high trading volume, likely contribute to the evidence that the trading of individuals is highly correlated and persistent. Grinblatt and Keloharju (2001) evidenced that past return, reference price effect, tax-loss selling, and the size of the holding period capital gain or loss etc. affect trading. Overconfident investors and sensation seeking investors trade more frequently (Grinblatt and Keloharju, 2009). Hong and Stein (2007) found that trading volume appears to be an indicator of sentiment. In other words, when prices look to be high relative to fundamental value, disagreement on price is strong, and volume is abnormally high. Thus, they proposed a disagreement volume model which allows speaking directly to a joint behavior between stock price and trading volume.

Whatever excessive trading (volume) hypotheses are overconfidence, sensation seeking, and disagreement, they all converge to a point that trading volume reflects the intensity of investors' emotion, belief, and preference.

Pavlov (1904), a Russian physiologist and Nobel laureate in physiology or medicine in 1904, proposed conditioned reflex (classical conditioning) when studied physiological reflex in animal, using saliva volume to measure conditioned reflex intensity in dog. Conditioned reflex is a physiological response to expect that an unconditioned stimulus will follow whenever a conditioned stimulus is present.

Thorndike (1913) is a pioneer in operant conditioning study. Skinner (1938) invented a conditioning chamber, called as a Skinner box, to study operant conditioning. It was a finding that a rat settles into a smooth pattern of frequent bar pressing, after it gets foods (reinforcement) several times by doing this. Today, psychologists define an operant reinforcement as any event that follows a response and increases its probability.

Like classical conditioning, operant learning is based on information and expectancies. Operant conditioning is a behavioral response with expectancy that if a certain operation is made, it will be followed by a certain consequence (reinforcement) at certain times. In a viewpoint, operant conditioning is that a discriminative stimulus sets the occasion for operant behavior, which is followed by a consequence (Pierce and Cheney, 2004). Or it occurs in the presence of certain stimuli and is always followed by certain consequences (Irons and Buskist, 2008). Skinner called this relationship a three-term contingency (Dragoi, 1997). In operant conditioning, reinforcement is used to alter the frequency of responses. It produces very high operant response rate and tremendous resistance to extinction that reinforcement follows the uncertain number of operant times (variable ratio) and time interval (variable interval) (Coon, 2007).

Intra-cranial stimulation (ICS) that involves direct stimulation of “pleasure centers” in brain is one of the most unusual and powerful reinforcement (Olds and Fobes, 1981). Some rats press a bar thousands of times per hour to obtain stimulation in experiment, ignoring food, water, and sex in favor of the bar pressing.

Coon (2007) classified operant reinforcement into three categories: primary reinforcement, secondary reinforcement, and feedback. Primary reinforcement is natural, or unlearned. They are usually rooted in biology and produce comfort, end discomfort, or fill an immediate physical need. Money, praise, approval, affection, and similar rewards, all serve as learned or secondary reinforcement. Secondary reinforcement that can be exchanged for primary reinforcement gain their value more directly. Printed money obviously has little or no value of its own, neither eaten nor drunk. However, it can be exchanged for foods and services, and perhaps is the most important source of economic reinforcement or conditioned reinforcement (Pierce and Cheney, 2004). For example, chimpanzees were taught to work for tokens in research (Cowles, 1937). Feedback is a process that an operator makes an (input) adjustment on his behavior after receiving the consequence of his response (output).

Soros (1995) studied reflex theory in a perspective of philosophy at beginning, and gradually found that it was correlated with stock price behavior. In his opinion, it is the best platform to study reflex in stock market. He has descriptive explanation for a feedback process, the third reinforcement in Coons’ classifications, that investors participate in trading, cause price volatility, make a judgment from price information, and then decide trading again to further influence price.

Studied joint behavior for security transaction volume distribution over a trading price range on every trading day, Shi (2006) derived a time-independent transaction volume-price probability wave equation and got two sets of analytical transaction volume distribution eigenfunctions over a trading price range in which transaction volume probability describes price volatility uncertainty and intensity using normative econophysics methodology. Based on this, we develop a theoretical trading conditioning model subject to price volatility and return information in terms of market psychological behavior, in which we use transaction volume probability to describe trading conditioning intensity. By studying correlation between the rate of return and that of change in trading conditioning intensity, we test both disposition effect and herd behavior simultaneously, trace back to analyze investors' psychological behavior with price volatility when they make trading decision, and attempt to provide a quantitative analysis and test model for behavioral finance.

Although there are many literatures on the relation between return and volume, we have not yet found study on the relation between the rate of return and that of change in transaction volume (probability), and a successful result that incorporates normative analytical transaction volume-price probability wave behavior in econophysics with descriptive cognitive reflex and conditioning behavior in physiology and psychology—it is difficulty for most of non-Expected Utility models in trying to achieve both normative and descriptive goals that they end up doing an unsatisfactory job at both (Barberis and Thaler, 2003). Unlike transaction volume, the rate of change in transaction volume (probability) could be positive, negative, or zero. Instead of testing disposition effect and herd behavior separately using event study and psychological experiment previously, moreover, we use high frequency data to test them simultaneously, and explain excessive trading and other anomalies in a unified theory.

The remainder of the paper is organized as follows. Section 2 will briefly introduce analytical transaction volume-price probability wave distributions and find a way to measure trading conditioning intensity subject to price volatility and return information; in section 3, we first study stationary equilibrium, determine the equilibrium price, and demonstrate the early finding that stationary equilibrium is prevailing in stock market (Shi, 2006) using Huaxia SSE 50ETF every trading high frequent data. Then, we empirically test correlations among the rate of mean return, the rate of change in trading conditioning intensity, and that of change in the amount of transaction in any two consecutive trading days. Section 4 is analyses and discussions on empirical results. They are: 1) Stationary equilibrium exists widely as a consequence of interaction and coherence in stock market; 2) General speaking, there are significant both disposition effect and herd behavior; 3) Cognitive and trading behavior changes with price volatility and environment; 4) Excessive trading explanation in trading conditioning theory; and 5) Potential application. Final are summaries and main conclusions.

## 2. TRANSACTION VOLUME DISTRIBITION AND TRADING CONDITIONING INTENSITY

### 2.1 Transaction Volume-price Probability Wave Distribution

Shi (2006) documented that stationary equilibrium exists widely in stock market by studying transaction volume-price behavior, using econophysics method. Transaction volume $v$ is accumulative trading volume at a price in a given time interval. It shows a kurtosis near the price mean value over a trading price range on every trading day regardless of stock price, price volatility path and direction[1]. A stationary equilibrium price is the price to which transaction volume kurtosis is corresponding. In addition, while trading price is volatile upward and downward to a stationary equilibrium price constantly, the equilibrium price jumps from time to time[1].

[1] Shi (2006) fitted and tested 618 volume distributions over a trading price range. Although there are different in price, price volatility path and direction, they show the same characteristic that accumulative trading volume exhibits kurtosis near the price mean value.

There are many factors that could influence stock price and its volatility, for example, management in listing company, macro-economy, news announcement, and psychological behavior etc. These factors impact on price change more and less if and only if there is trading. Shi (2006) did not consider them temporarily in his previous study.

There is no price volatility at all if there is no trading in stock market. But if there is trading, we can not draw a conclusion that there is price change. Trading is a necessary condition for price volatility. In addition, the amount of transaction constrains price volatility and transaction volume.

Studied how the amount of transaction constrains price volatility and transaction volume, and how the volume distributes over a trading price range in stationary equilibrium in stock market, Shi (2006) proposed a relationship hypothesis among the constraint of transaction amount, the transaction volume distribution of the amount, and the price volatility of the amount (price reversal to a stationary equilibrium price), i.e. transaction amount constraint hypothesis:

$$-E + p\frac{v_t^2}{V} + W(p) = 0, \qquad (1)$$

where $p$ is price, its dimension is [currency unit][share]$^{-1}$ ; $v_t = v/t$ is transaction momentum and transaction volume in a given time interval $t$, its dimension is [share][time]$^{-1}$; $E = p \cdot v/t^2$ is the amount of transaction at a correspondent price $p$ in a given time interval $t$, its dimension is[currency][time]$^{-2}$; $v_{tt} = v/t^2$ is impulse and momentum force in a direction deviating from a stationary equilibrium price at a corresponding price $p$ in a given time interval $t$, its dimension is[share][time]$^{-2}$, $p \cdot v_t^2 / V$ is the volume probability of transaction amount at a price $p$ in a given time interval $t$, its dimension is [currency unit][time]$^{-2}$; In addition,

$$W(p) = A(p - p_0) \approx A(p - \overline{p}) \qquad (2)$$

is the price volatility and linear reversal of transaction amount to a stationary equilibrium price at a price $p$ in a given time interval $t$, its dimension is [currency unit][time]$^{-2}$, too; $p_0$ is a stationary equilibrium price and $|\overline{p}|$ is price mean value; $A = (1 - v/V)v_{tt}$ is the magnitude of a supply-demand restoring force or stationary equilibrium restoring force , its dimension is [share][time]$^{-2}$.

According to stock trading regulation, transaction priority is given by “price first and time first”. If current trading price is $p_c$ , then next trading priority is given to minimize price volatility with respect to it. It states that actual trading price path is chosen by transaction amount constraint hypothesis, equation (1), functional

$$F(p,\psi)=(W-E)\psi^*\psi+\frac{B^2}{V}p\left(\frac{d\psi^*}{dp}\right)\left(\frac{d\psi}{dp}\right) \tag{3}$$

to minimize its wave function $\psi(p)$ with respect to price variations. Its mathematical expression is

$$\delta\int F(p,\psi)dp=0. \tag{4}$$

[1] Unlike Poisson-diffusion jump (Ait-Sahalia, 2004) and Levy jump (Li et al., 2008), stationary equilibrium price jump has its clear mechanism. For example, if there is sudden increasing buying volume to break stationary equilibrium, trading price will be volatile upward and downward from a stationary equilibrium price to another by its jump. We can measure its jump behavior in terms of volume distribution over a trading price range (Shi, 2006).

From equation (1) and (4), we have transaction volume-price probability wave equation

$$\frac{B^2}{V}\left(p\frac{d^2\psi}{dp^2}+\frac{d\psi}{dp}\right)+[E-W(p)]\psi=0. \tag{5}$$

Substituting equation (2) into equation (5) with its natural boundary conditions

$\psi(0)=0$, $\psi(p_0)<\infty$, and $\psi(+\infty)\to 0$,

we get two sets of analytical transaction volume distribution eigenfunctions over a trading price range. One is that when there is coherence (the sum of momentum force $v_{tt}$ and restoring force $A$ is equal to an eigenvalue constant over a trading price range), we have

$$\psi_m(p)=C_m J_0[\omega_m(p-p_0)], \qquad (m=0,1,2,\cdots) \tag{6}$$

where $\omega_m$ satisfy

$$\omega_m^2=v_{tt}-A=\frac{v}{V}v_{tt}, \qquad (\omega_m>0) \qquad (m=0,1,2,\cdots) \tag{7}$$

a set of positive eigenvalue constants, $J_0[\omega_m(p-p_0)]$ are a set of zero-order Bessel eigenfunctions, $C_m$ are normalized dimensionless constants; the other is that when restoring force $A_m$ are a set of constants over a trading price range, we get the other set of multi-order eigenfunctions as

$$\psi_m(p)=C_m e^{-\sqrt{A_m}|p-p_0|}\cdot F\left(-n,1,2\sqrt{A_m}|p-p_0|\right), \quad (n,m=0,1,2,\cdots) \tag{8}$$

where $$\sqrt{A_m}=\frac{E_m}{1+2n}=const.>0, \qquad (n,m=0,1,2,\cdots) \tag{9}$$

$F\left(-n,1,2\sqrt{A_m}|p-p_0|\right)$ are a set of $n$ order confluent hypergeometric eigenfunctions or the first Kummer's eigenfunctions.

The absolute of functions (6) and (8) $|\psi_m(p)|$ is transaction volume density and probability at price $p$ in a given time interval (see figure 1 and figure 2, in which x-coordinate is price,

y-coordinate is transaction volume distribution, and origin is a stationary equilibrium price), $C_m$ is normalized constant.

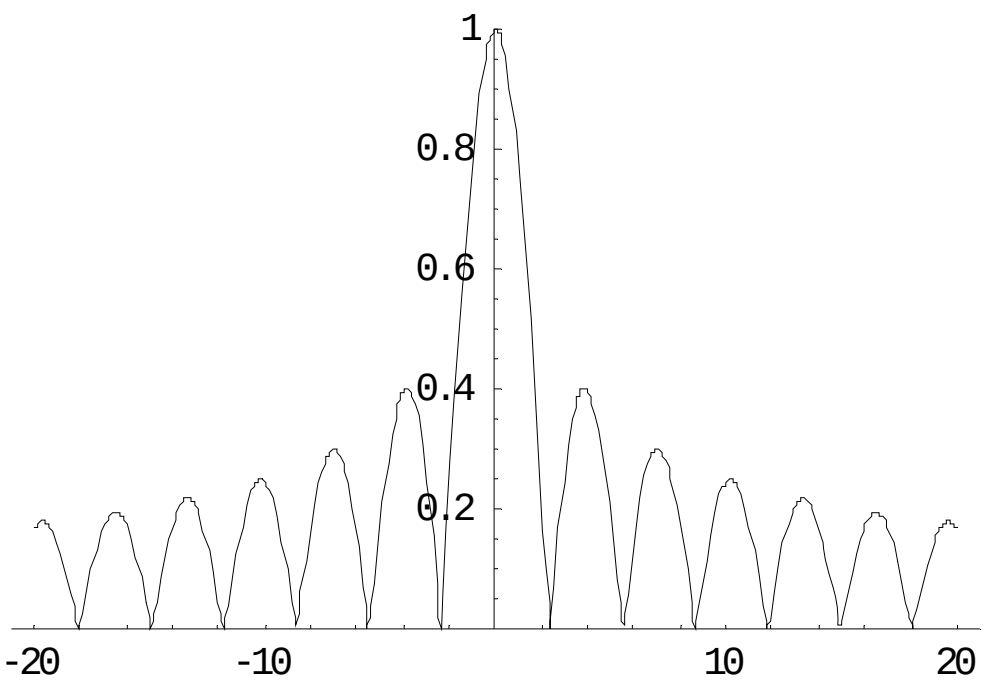

Figure 1: Transaction volume distribution over a trading price range in the absolute of function (6)

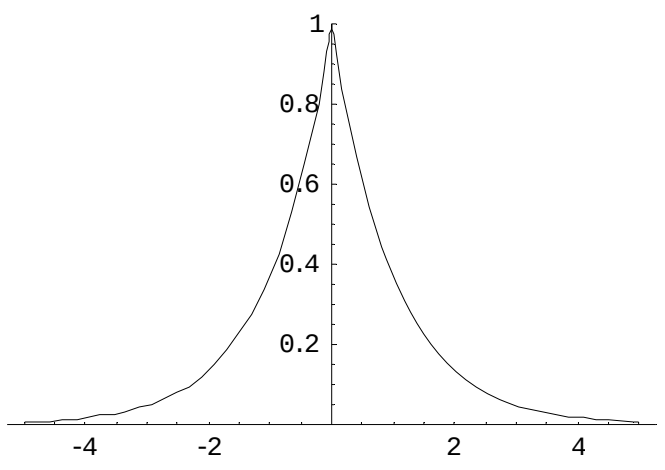

(a) Zero-order

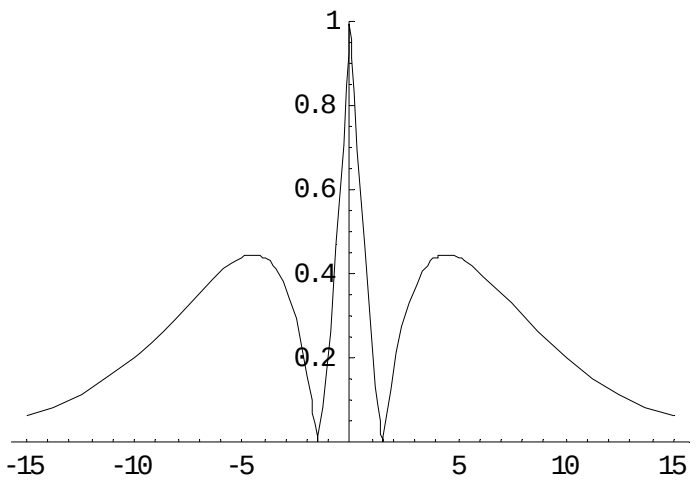

(b) The first-order

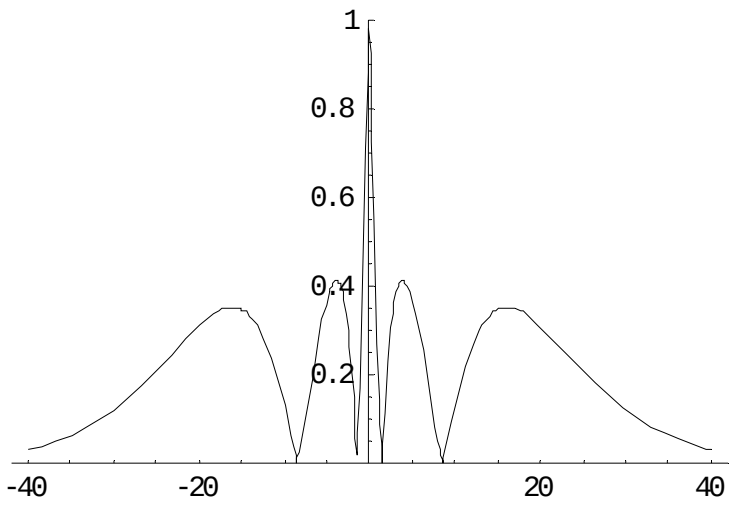

(c) The second-order

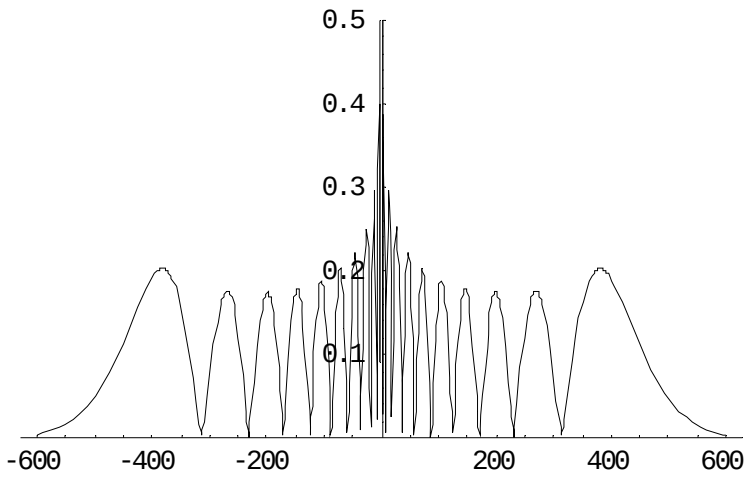

(d) the tenth-order

Figure 2: Transaction volume distribution over a trading price range in the absolute of function (8)

There are several different characteristics between figure 1 and figure 2. First, the volume distribution in figure 1 is an attenuated wave with price departing from stationary equilibrium price, while it is a uniform wave with price except for zero-order distribution which is exponent in figure 2. The higher the order is in the volume distribution function, the better uniformly the volume distributes over a price range in figure 2. Second, it is an open distribution in figure 1, which can describe a pulse trading behavior far away from a stationary equilibrium price, for example, profit and loss transfer trading, whereas it is a closed distribution in figure 2, which can describe random and uniform behavior. Third, the magnitude of eigenvalue in figure 2 is about two orders of magnitude larger than that in figure 1 except for exponent distribution. However, both distributions can fit exponent distribution very well in common.

## 2.2 Eigenvalue and Probability Wave

An eigenvalue is different from a parameter in a distribution function when we use it to describe uncertainty in events. In a great majority of scenarios, we usually use a distribution function with a parameter to analyze statistic and uncertain behavior in events if we know their distribution in a certain extent, but do not understand its mechanism. In this approximate analysis, we do not clear what exact information is contained in the parameter. An eigenvalue is observable and measurable. Thus, it is possible for us to test distribution eigenfunction validity quantitatively and understand its mechanism. In transaction volume-price probability wave distribution, for example, two sets of eigenvalues represent two observables. One is calculated by equation (7). It is the consequence of a kind of coherence that the sum of momentum force $v_{tt}$ and supply-demand restoring force is equal to this eigenvalue constant over a trading price range. The other is calculated by equation (9). It is the consequence that supply-demand restoring and reversal force is equal to the eigenvalue constant over a trading price range. It can describe exponent distribution and uniform random distribution.

(a) Classical wave (b) Probability wave

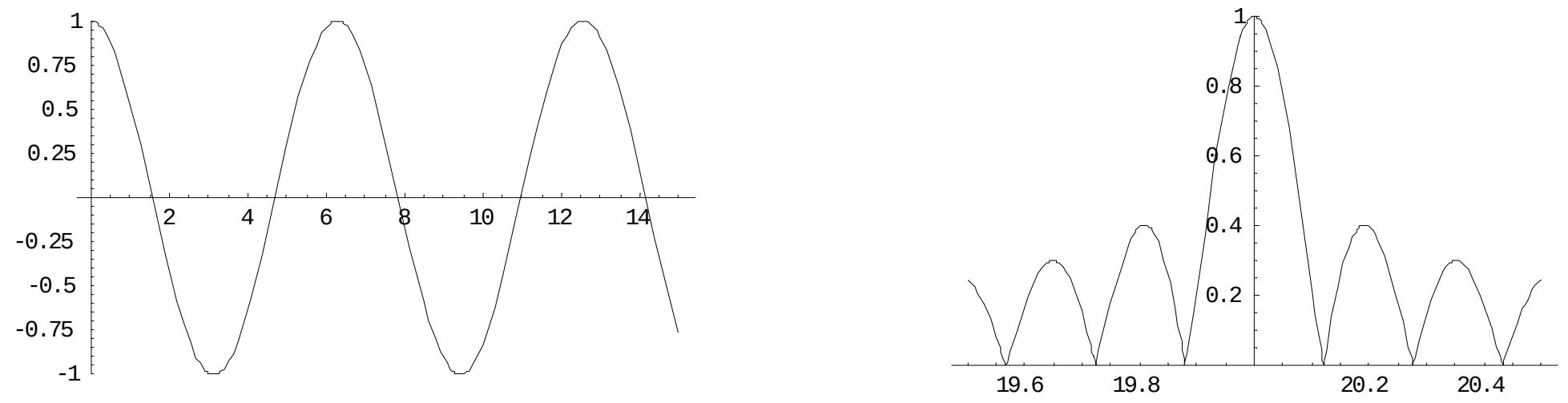

$$y(t) = y_0 \cos(\omega t)$$

$$|\psi_m(p)| = C_m |J_0[\omega_m(p - p_0)]|$$

Figure 3: Classical wave and probability wave

Probability wave is a kind of wave in which we use volume distribution and probability rather than the amplitude of wave to describe wave intensity. For example, we use transaction volume probability rather than the amplitude of price volatility to describe price volatility intensity in a transaction volume-price probability wave. Now, let us contrast and compare probability wave with classical wave. First, x-coordinate is price (in stock market) or distance (in quantum mechanics) and y-coordinate is accumulative volume probability in a probability wave, while x-coordinate is time and y-coordinate is amplitude in a classical wave. Second, we use volume distribution and probability to describe wave intensity, whereas we use the amplitude of wave to describe wave intensity. Third, intensity is equal to and larger than zero in probability wave, while it can be negative in classical wave; Forth, intensity displays wave change with an independent variable in a probability wave. Large amplitude is not equal to strong intensity in a probability wave. In a classical wave, the larger the amplitude is, the stronger the intensity does be in wave. Fifth, probability wave is the consequence of coherence in group. It can not exist independently. Classical wave can exist independently. Sixth, strictly speaking, we have not yet found that there is a periodicity and time cycle in a probability wave. It is that there is uncertainty in time prediction. In classical wave, we can measure speed in wave. Thus, there is periodicity and time cycle. However, there are common in both. There is coherence in both probability wave and classical wave. And there is repeated change reference to an equilibrium point (see figure 3).

## 2.3 Price Volatile Trading Conditioning and Its Intensity

Let us first consider a rigid trading system in which there is no any subjective cognition change in buying and selling behavior. The transaction volume is the same at any the same time interval even if trading price is volatile.

If there is incremental amount of transaction to buy, to break stationary equilibrium, and to cause stationary equilibrium price jump, then, there is one-to-one correspondence between incremental amount of transaction and the equilibrium price jump because total transaction volume is the same in two time intervals before and after jump.

In a real financial market, there are a great many factors that could influence investors' subjective cognition, supply-demand balance, and transaction volume. Therefore, it is a key to find a certain relation among incremental amount of transaction, transaction volume change, and stationary equilibrium price jump how we combine market subjective trading behavior and transaction volume change into transaction volume-price probability wave distribution function. It is also a key to develop a theoretical trading conditioning model subject to price volatility and return information.

It is objective that people have physiological demand for clothes, foods, and services in life. This is physiological response and unconditioned reflex. Printed money obviously has little or no value of its own, neither eaten nor drunk. In commodity exchange economy, however, when we associate money, income, and return with the necessities of life and services tightly through exchange, we have been conditioned by them (conditioned reinforcement) and will produce the same physiological response to them, i.e. conditioned reflex. Pierce and Cheney (2004) wrote that money can be exchanged for foods and services, and perhaps is the most important source of economic and conditioned reinforcement.

In stock market, it is a conditioned reflex (classical conditioning) that traders produce the same physiological response when they are stimulated by price volatility and return information as they do for the necessities of life and services, and it is a trading conditioning that practitioners trade and expect return in the future after they analyze, judge and have decision making in the presence of price volatility and return information.

The reinforcement is money and return in price volatile trading conditioning. It has several characteristics as follows: First, it does not loss reinforcement value as quickly as primary reinforcement does because physiological demand is satisfied. Second, profit is a positive reinforcement whereas loss is a negative reinforcement. People trade and buy if they expect price rising. On the other hand, they trade and sell if they expect price dropping. There are tradings no matter whether price rises or drops in expectancy. Third, because we have not yet found time cycle in volume-price probability wave, return occurs after stock holding in an uncertain time interval. This makes tremendous resistance to extinction in trading conditioning, similar to variable interval (VI) schedules in operant conditioning experiment (Coon, 2007). Forth, the trading consequence (profit and loss) that produces feedback to traders could influence their emotion and judgment, and enhance or change their expectancy in return to make trading. For examples, loss stop selling investors are likely to buy stock again because they change their expectancy in return, stimulated by price volatility and return. Just buying traders may sell their stock immediately because they find market movement in an opposite direction and change their previous expectancy. Therefore, cognitive response in price volatile trading conditioning includes both expectancy and expectancy change in return.

Based on Skinners' three-term contingency and operant conditioning (Irons and Buskist, 2008), we define price volatile trading conditioning in stock market as participants trade (operant) and expect return (profit and loss) in the future (consequence) after they analyze, judge, and have decision making (cognition) in the presence of price volatility and return information (discriminative stimuli). The trading consequence (profit and loss) that produces feedback to influences their emotion and judgment, adjust their expectancy in return, and make trading (see figure 4).

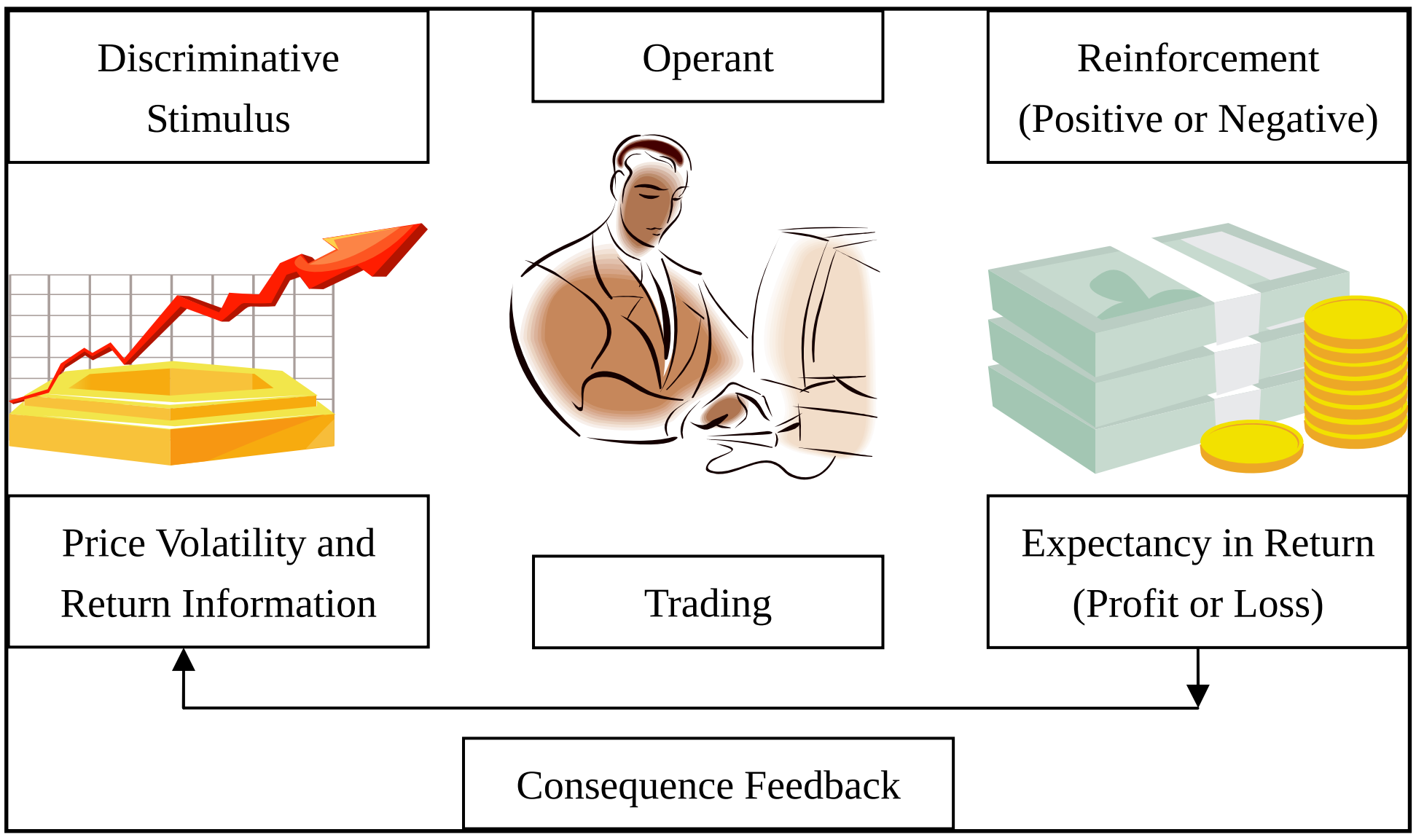

Figure 4: Three-term contingency in trading conditioning

There are two independent variables, price and transaction volume, in stock market. The amount of transaction is a constrain condition on them. According to classical mechanics (Greenwood, 1977), the degree of freedom is equal to that the number of independent variables minus the number of constrain conditions. The degree of freedom is one in this trading system.

Osborne (1977), a pioneer in econophysics, found that price as a function of volume does not exist empirically, and then explained why volume is the function of price, and this is not invertible. McCauley (2000) proved that price as a function of volume does not exist mathematically. Therefore, Shi (2006) chose price as an independent variable and transaction volume as a dependent variable in stock market.

According to afore-developed trading conditioning behavior, security price volatility and return information stimulate practitioners to trade with expectancy on its return in the future. The subjective trading behavior (cognition) and accumulative trading volume (transaction volume) change synchronously. We use the change in transaction volume distribution (probability) over a trading price range to present the change in subjective trading behavior (cognition) subject to price volatility and return information (We use transaction volume rather than the amount of transaction because the later is constraint).

In transaction volume-price probability wave distribution function, we use transaction volume probability instead of the amplitude of price volatility to describe price volatility intensity. The larger the transaction volume probability is at a price, the higher the trading intensity is (Unlike a classic wave in that the larger the amplitude is, the stronger the intensity does be). The higher the trading intensity is, the higher the trading frequency[1]. Therefore, expectancy on return and trading conditioning intensity are higher (The higher the transaction volume is, the higher volume traders buy. There are more expectancy on price rising. On the other side, the higher the transaction volume is, the higher volume traders sell. There are more expectancy on price dropping), and vice versa. Price volatile intensity is approximately equal to trading conditioning intensity subject to price volatility and return information. Therefore, we can use transaction volume probability (distribution) to describe trading conditioning intensity and frequency (reference to figure 5).

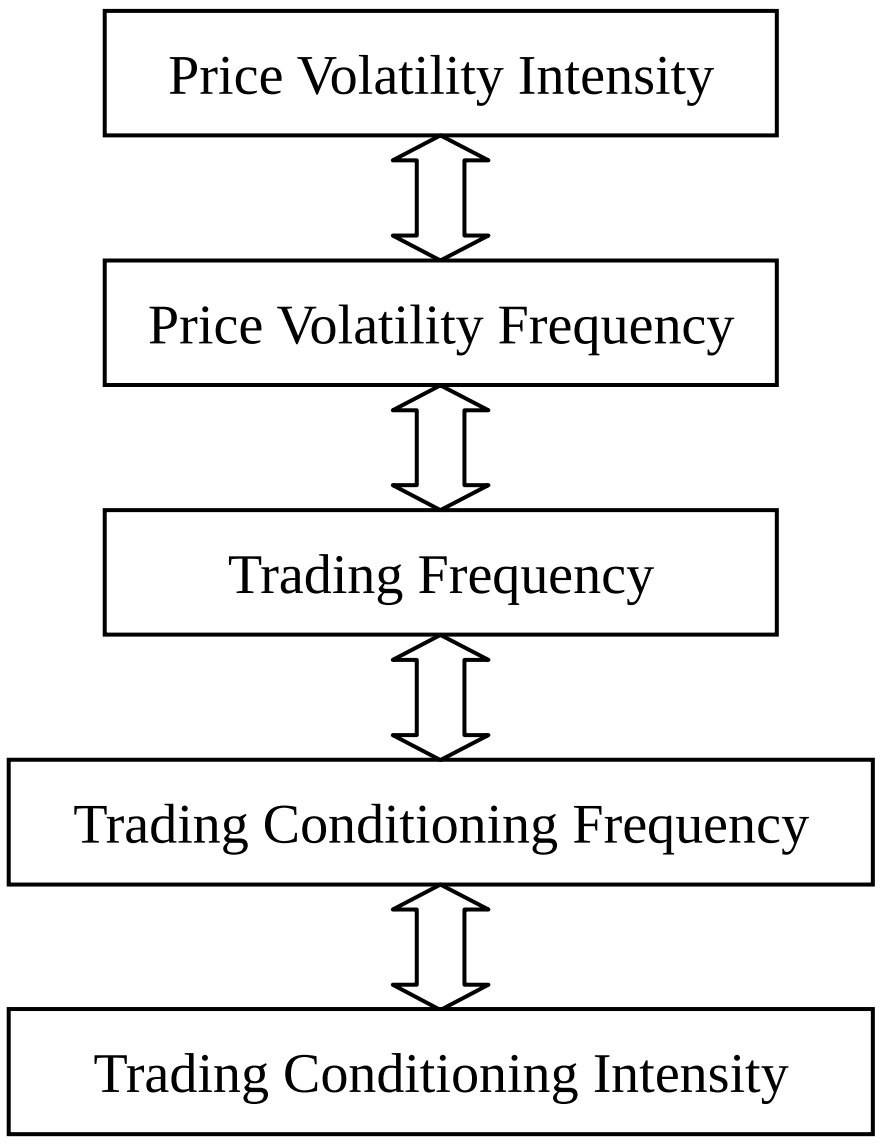

Figure 5: Price volatility intensity is approximately equal to trading conditioning intensity

Now, we can use transaction volume probability to describe trading conditioning intensity. In the same way, when we use stationary equilibrium jump to represent price volatility mean return in two consecutive trading days, we can use the rate of change in total transaction volume to describe the rate of change in trading conditioning intensity subject to the rate of mean return and its information in the two days. Obviously, the rate of change in trading conditioning intensity could be positive, negative, and zero. In this way, we can study correlation between the rate of mean return and that of change in trading conditioning intensity and trace back to analyze investors' physiological response and psychological behavior when they trade in decision making, using high frequency data.

[1]According to the law of large number in probability and statistics, if total transaction volume is much greater than the volume of every trading, then, transaction volume probability is approximately equal to trading frequency. In our study, total daily trading volume is about 360,000,000 shares in average. We can use transaction volume probability to represent trading frequency. The higher the transaction volume is at a price, the higher trading frequency. Although there is difference in information contained in a single large trading volume equivalent to total volume of several small trades, transaction volume probability is still approximately equal to trading frequency. The abnormal volume distribution disturbed by large volume trading reveals that stationary equilibrium is easily broken by capital advantage speculators. We will study it in the future.

## 3. EMPIRCAL TESTS

In this section, we will fit and test accumulative trading volume distribution over a trading price range by using the absolute of zero-order Bessel eigenfunction regression model. From this, we can get a stationary equilibrium price on every trading day and calculate its jump and mean return in any two consecutive trading days. For those that show the significance fitted by the model, we can determine a stationary equilibrium from fitting figures directly. Otherwise, we substitute price mean value for it. In this way, we can further study the correlations among the rate of mean return, the rate of change in trading conditioning intensity, and that of change in the amount of transaction.

### 3.1 Data

We use every trading high frequent data in Huaxia SSE 50ETF from April 2, 2007 to April 10, 2009, in which there are nearly 740 days and 495 trading days in total, i.e., the total number of volume distributions over a trading price range is 495. The data is from HF2 database, provided by Harvest Fund Management Co., Ltd.

We process the data in two steps. First, we reserved two places of decimals in price by rounding-off method and added volume at a corresponding price (original data reserves three places of decimals.). Second, transaction volume at a price is divided by total transaction volume over a trading price range. Thus, we acquire transaction volume probability (distribution) over a trading price range in every trading day.

### 3.2 Distribution Fitting, Significance Test, and Stationary Equilibrium Price

In a stationary equilibrium state, theoretical transaction volume-price distribution function is

$$|\psi_m(p)| = C_m |J_0[\omega_m(p-p_0)]|, \qquad (m=0,1,2\cdots) \qquad (10)$$

where $C_m$, $\omega_m$ and $p_0$ are a normalized constant, an eigenvalue constant and a stationary equilibrium price, respectively, and determined by a nonlinear regression model as follow:

$$|\psi_m(p_i)| = C_m |J_0[\omega_m(p_i-p_0)]| + \varepsilon_i, \qquad (i=1,2,3\cdots,n) \qquad (11)$$

where $n$ is the number of prices over a trading price range in a trading day; $\varepsilon_i$ is random error subject to $N(0,\sigma^2)$; $|\psi_m(p_i)|$ and $C_m|J_0[\omega_m(p_i-p_0)]|$ are observed and theoretical transaction volume probability at a price over a trading price range, respectively. We fit the volume distribution by Levenberg-Marquardt nonlinear least square method, get the numerical values of $C_m$, $\omega_m$ and $p_0$, and find theoretical distribution. Origin 6.0 Professional software is friendly used in fitting (see figure 6 (a)).

We use $F$ statistic to test significance. The coefficient of determination $R^2$ is as follow:

$$R^2 = \frac{ESS}{TSS} = \frac{TSS - RSS}{TSS}, \qquad (12)$$

where $ESS = \sum_{i=1}^{n}\left(\hat{Y}_i - \overline{Y}\right)^2$, $RSS = \sum_{i=1}^{n}\left(Y_i - \hat{Y}_i\right)^2$, and $TSS = \sum_{i=1}^{n}\left(Y_i - \overline{Y}\right)^2$ are the explained sum of squares, the residual sum of squares, and total sum of squares, respectively. And,

$$F = \frac{ESS / k}{RSS / (n - k - 1)} \qquad (13)$$

where $n$ and $k$ is sample size and the number of explanatory variables, respectively. If $F > F_{0.05}$ or

$$R^2 > R_{crit}^2 = \frac{k \cdot F_{0.05}}{k \cdot F_{0.05} + (n - k - 1)}, \qquad (14)$$

the regression model (11) holds true at 95% significant level.

Our test results show that 380 out of 495 distributions show significance from April 2, 2007 to April 10, 2009 (76.77%). The remainders (23.23%) lack significance.

There are two notable characteristics among significance lacking distributions. First, the number of trading prices is few or the sample size is too small in price. It is partly credited by previous data process, in which we reserved two places of decimals in price by rounding-off three places in original data. The volume distribution characteristics can not be displayed.

To solve the problem, we add 0.005 in three places of decimals in price and subdivided volume at corresponding prices. As a result, 28 volume distributions show significance, modeled by absolute zero-order Bessel function. Thus, there are total 408 (380+28) volume distributions, around 82.42%, that show significance. Their stationary equilibrium prices can be determined by fitting results directly.

Second, the remaining volume distributions (87) show at least two of kurtosis over a trading price range. It is explained when abrupt change in supply and demand quantity, for example, continuous large buying volume, breaks up original stationary equilibrium, price volatility is going to be adjusted to a new equilibrium price. The stationary equilibrium price appears a step and jump change. After this, trading price is volatile around the new stationary equilibrium price. In this case, the volume distribution function is the linear superposition of function (10), that is,

$$\left|\psi_m(p)\right| = \sum_n C_m \left|J_0\left[\omega_{m,n}(p - p_{0n})\right]\right|, \qquad (n = 1,2\cdots) \qquad (15)$$

where $n$ is the number of stationary equilibrium prices. We fit them with a two stationary equilibrium price regression model as follow:

$$\left|\psi_m(p_i)\right| = \sum_i \sum_{n=1,2} C_m \left|J_0\left[\omega_{m,n}(p_i - p_{0,n})\right]\right| + \varepsilon_i \qquad (i = 1,2\cdots) \qquad (16)$$

where $n = 2$.

We test significance ($R_2^2 > R_{2crit}^2$, here $k = 2$). Of 87 distributions, 59 distributions (11.92%

in total) show significance at 95% level (see figure 6 (b)).

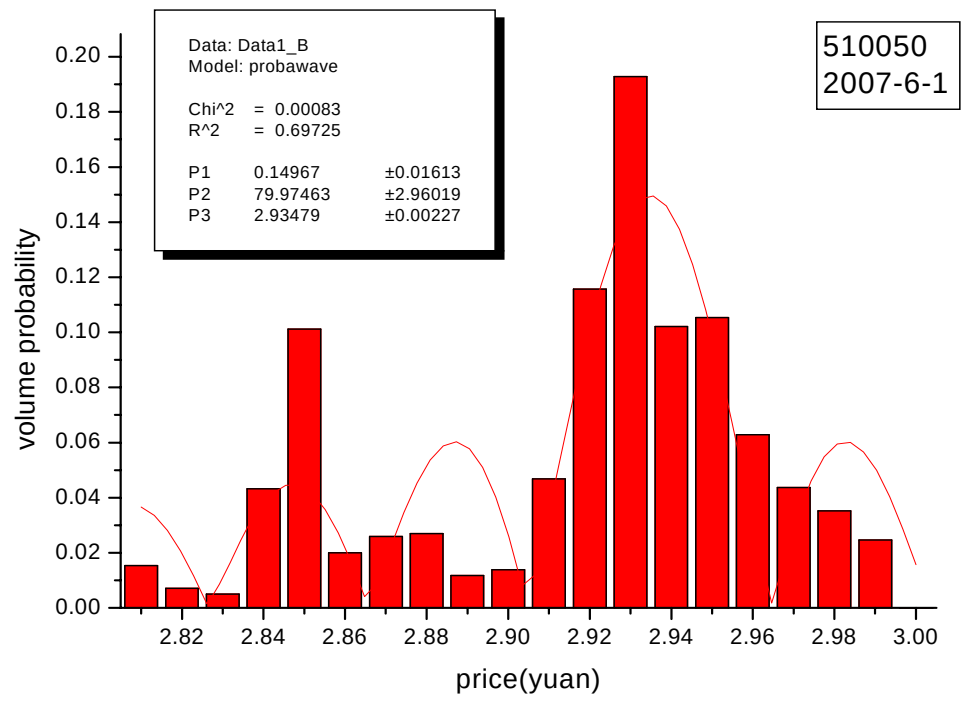

(a) A typical sample is fitted by function (10)

( $R^2 = 0.697 > R^2_{crit} = 0.50$ )

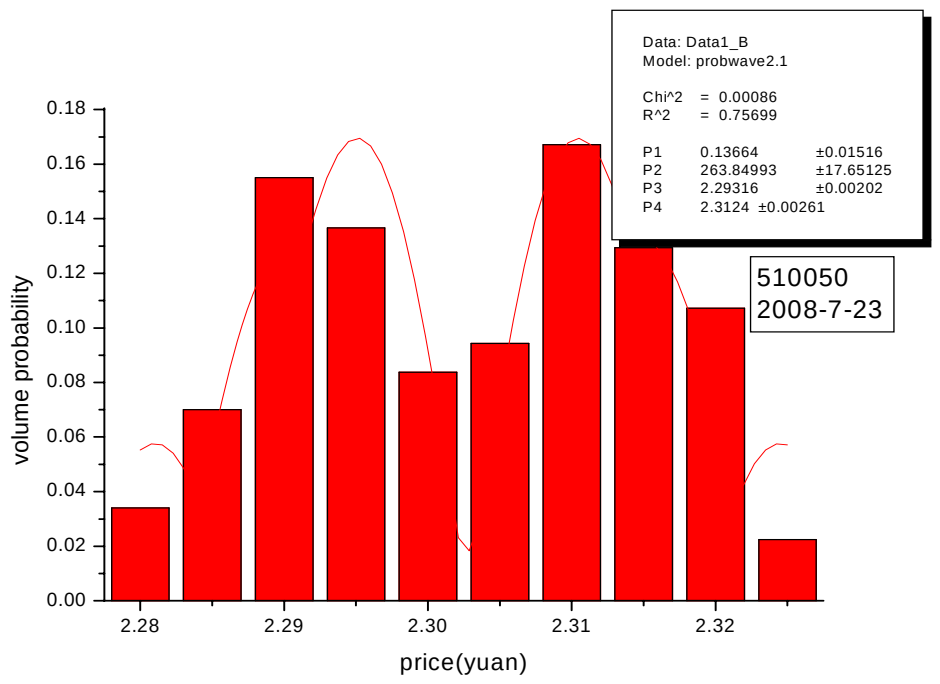

(b) A typical sample is fitted by function (16)

( $R_2^2 = 0.757 > R^2_{2crit} = 0.575$ )

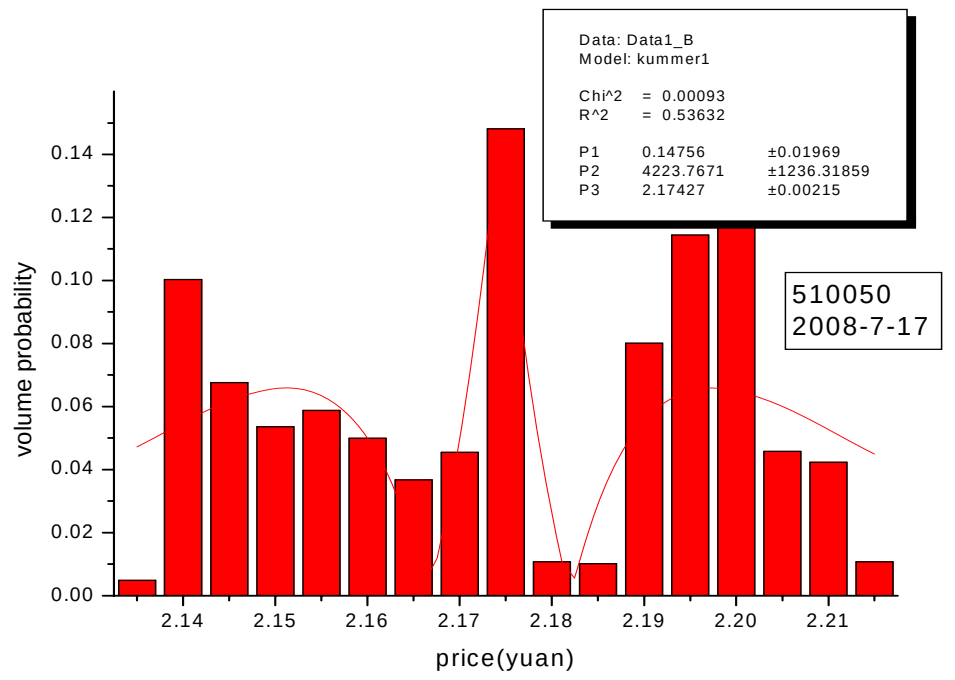

(c) A typical sample is fitted by function

( $R^2 = 0.536 > R^2_{crit} = 0.232$ )

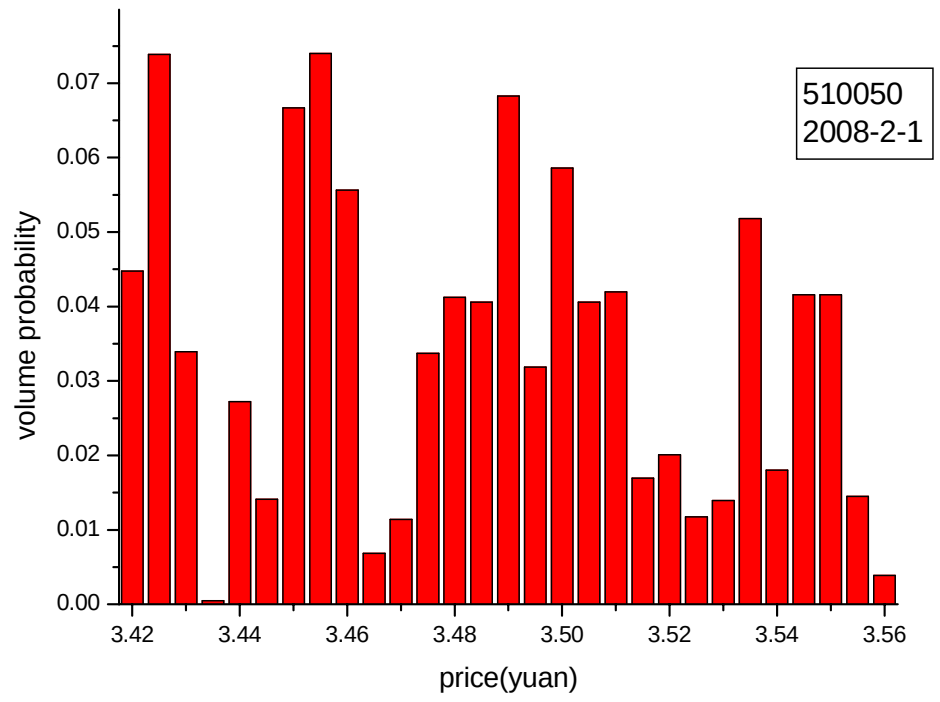

(d) One of five insignificant samples

Figure 6: Distributions fitted by transaction volume-price probability wave model and significance test [1]

For the rest of 28 transaction volume distributions, we fit them by the second set of distribution functions as follow:

$$\left|\psi_m(p)\right| = C_m e^{-\sqrt{A_m}|p-p_0|} \cdot \left|F\left(-n,1,2\sqrt{A_m}|p-p_0|\right)\right| . \qquad (17)$$

In convenience, we choose $n = 1$, it is

$$\left|\psi_m(p)\right| = C_m e^{-\sqrt{A_m}|p-p_0|} \cdot \left|F\left(-1,1,2\sqrt{A_m}|p-p_0|\right)\right|$$

$$= C_m e^{-\sqrt{A_m}|p-p_0|} \cdot \left|1 - 2\sqrt{A_m}|p-p_0|\right| . \qquad (18)$$

The result is that 23 distributions (about 4.65% in total) show significance at 95% level by this regression model (see figure 6 (c)). The last 5 distributions lack significance (see figure 6 (d)).

[1] In figure, P1, P2, and P3 are a normalized constant, an eigenvalue, and a stationary equilibrium price, respectively. P4 is a stationary equilibrium price, too.

In figure 6 shows some typical test results fitted by transaction volume-price distribution functions (10), (16), and (18), respectively.

### 3.3 Correlation and Significant Test

408 transaction volume distributions (about 82.42% in total samples) show significance, tested by the absolute of zero-order Bessel eigenfunction regression model. We can decide their stationary equilibrium prices from fitting figures directly. For the rest samples, we substitute price mean values for them. In this way, we can figure out price volatility mean return by stationary equilibrium price jump in any two consecutive trading days. Thus, we can study the correlation among the rate of mean return, the rate of change in trading conditioning intensity, and that of change in the amount of transaction. Here, the rate of change in trading conditioning intensity is approximately equal to that in total transaction volume in any two consecutive trading days.

Given correlation coefficient $r_{X,Y}$ as

$$r_{X,Y} = \frac{\mathrm{cov}(X,Y)}{\sigma_X \sigma_Y}, \tag{19}$$

where $\sigma_X$ and $\sigma_Y$ are the standard deviations of variable $X$ and $Y$, $\mathrm{cov}(X,Y)$is covariance. We use $t$-statistic to test significance. If we have

$$H_0 : \rho = 0 \quad \text{and} \quad H_1 : \rho \neq 0,$$

then,
$$t = \frac{|r - \rho|}{\sqrt{(1 - r^2)/(n - 2)}}, \tag{20}$$

where $r$ and $n$ are correlation coefficient and sample size, respectively. For $\alpha = 0.05$, if $t > t_{crit} = t_{0.05/2}(n-2)$, then, original hypothesis is rejected. Correlation coefficient is significant not equal to zero at 95% level.

In our test, we use Eviews6.0. We subdivide our data into 5 groups in terms of time, financial crisis in the United States, and bubble crash in China. The first group is dated from April 2, 2007 to June 29, 2007, the first half before bubble crash in China. The second group is dated from July 2, 2007 to October 31, 2007, the second half before bubble crash in China. The third group is dated from November 1, 2007 to April 40, 2008, the first half after bubble crash in China. The forth group is dated from May 5, 2008 to October 31, 2008, the second half after bubble crash in China. And latest one is dated from November 3, 2008 to April 10, 2009, the reversal time after a year time deep dropping (reference to Table).

There is a major advantage when we subdivide data into 5 time intervals. We can find how market psychological behavior change with time and environment and quire why.

By empiric results, we have main findings as follows: generally speaking, first, there are significant positive correlations between the rate of mean return and that of change in both trading conditioning intensity and the amount of transaction; second, correlation coefficient varies in 5 subdivided periods. For examples, (a) they lack significance in spite of positive correlations in two time intervals right before and just after bubble crashes; (b) it shows positive significance in the second half after bubble crash; (c) the positive correlation coefficient is 0.4766, the highest during reversal period after a year time deep dropping; (d) particularly, there exists significant negative correlation between the rate of mean return and that of change in trading conditioning intensity when SEE Composite Index is rising during bull market; third, the correlation coefficient between the rate of mean return and that of change in amount of transaction is always several points (0.03~0.08) higher than that between the rate of mean return and that of change in trading conditioning intensity. We will discuss them in next section.

Table: Correlation and Significance Test

| | Time (yr. mo. d.) | Number of Distributions | SSE Composite Index | Correlation 1 | Correlation 2 | Differences | Correlation 3 |
|---|---|---|---|---|---|---|---|
| A | 2007.4.2—2009.4.10 | 494 | 3252.59—2444.23 | 0.1391 (t=3.115> $t_{crit}$ =1.960) | 0.1970 (t=4.458> $t_{crit}$ =1.960) | 0.0579 | 0.9975 (t=316.3> $t_{crit}$ =1.960) |
| B | 2007.4.2—2007.6.29 | 59 | 3252.59—3820.70 | -0.2567 (t=2.006> $t_{crit}$ =2.001) | **-0.2120 (t=1.638< $t_{crit}$ =2.001)** | 0.0447 | 0.9986 (t=142.0> $t_{crit}$ =2.001) |
| C | 2007.7.2—2007.10.30 | 83 | 3836.29—5954.77 | **0.0729 (t=0.6583< $t_{crit}$ =1.990)** | **0.1053 (t=0.9529< $t_{crit}$ =1.990)** | 0.0324 | 0.9993 (t=241.2 > $t_{crit}$ =1.990) |
| D | 2007.11.1—2008.4.30 | 122 | 5914.28—3693.11 | **0.1026 (t=1.130< $t_{crit}$ =1.980)** | **0.1714 (t=1.906< $t_{crit}$ =1.980)** | 0.0688 | 0.9968 (t=137.0> $t_{crit}$ =1.980) |
| E | 2008.5.5—2008.10.31 | 123 | 3761.01—1728.79 | 0.1963 (t=2.202> $t_{crit}$ =1.980) | 0.2706 (t=3.091> $t_{crit}$ =1.980) | 0.0743 | 0.9958 (t=119.2> $t_{crit}$ =1.980) |
| F | 2008.11.3—2009.4.10 | 107 | 1719.77—2444.23 | 0.4766 (t=5.556> $t_{crit}$ =1.983) | 0.5203 (t=6.243> $t_{crit}$ =1.983) | 0.0437 | 0.9981 (t=166.5 > $t_{crit}$ =1.983) |

Notes: (1) Correlation 1 is the correlation between the rate of mean return and that of change in trading conditioning intensity; correlation 2 is the correlation between the rate of mean return and that of change in the amount of transaction; Correlation 3 is the correlation between the rate of change in trading conditioning intensity and that in the amount of transaction; (2) Differences are correlation coefficient 2 minus 1; (3) $t_{crit}$ is $t_{0.05/2}(n-2)$; If $t > t_{crit}$, then, the correlation coefficient is significantly not equal to zero; On the other hand, we can not reject original hypothesis that the correlation coefficient is equal to zero; (4) the lack of significance is printed in bold and red; (5) SSE Composite Index is measured by closing point.

## 4. ANALYSES AND DISCUSSIONS ON EMPIRICAL RESULTS

### 4.1 Stationary Equilibrium and the Equilibrium Price

Shi (2006) decomposed price volatility behavior into two parts. First, trading price is volatile upward and downward to a stationary equilibrium price constantly. It is quantitatively described by a transaction volume-price probability wave equation (5), in which a stationary equilibrium price is the price that transaction volume kurtosis corresponds to. Second, a stationary equilibrium price jumps from time to time. It is explained that a stationary equilibrium is easily broken and its restoring force is weak in stock market. If there is abrupt change in supply and demand quantity, for example, by a large buying volume, then trading price is going to adjust to a new stationary equilibrium price that jumps. After this, it is volatile upward and downward to new equilibrium price. The jump is used to represent for mean return.

We use every trading high frequent data in Huaxia SSE 50ETF from April 2, 2007 to April 10, 2009, in which there are nearly 740 days and 495 trading days in total, i.e., the number of volume distribution over a price range is 495. We firstly apply transaction volume-price probability wave distribution regression model, equation (11), to fitting them, and test significance. 408 distributions or 82.42% in total show significance. For those distributions that show significance, we get stationary equilibrium prices directly from fitting results. For those that do not show significance, the volume distributions display two and more than two of kurtosis over trading price range in a day. It is that a stationary equilibrium price jumps on trading days. We substitute price mean value for stationary equilibrium price. Thus, we can get price volatility mean return in any two consecutive trading days by stationary equilibrium price jump.

Our empirical results further demonstrate the early finding that there is stationary equilibrium and a stationary equilibrium price jumps from time to time in stock market (Shi, 2006).

The ratio of the number of abnormal distributions to total number of distributions is 23.23% from 2007 to 2009 in this paper and 5.66% in 2003 (Shi, 2006). By comparison, we can draw a conclusion that it is much more stable and less volatile in Shanghai stock market in 2003. It is fact that there was 26.17% maximum volatility in 2003, whereas there was 140.96%, 231.71%, and 88.60% from 2007 to 2009, respectively. The larger the ratio is, the more volatile and unstable market is. Thus, the ratio can be used as a stability index in stock market.

### 4.2 Test Disposition Effect and Herd Behavior

From empirical test results in Section 3, we find that there is significant positive correlation between the rate of mean return and that of change in trading conditioning intensity in general from April, 2007 to April, 2009 (reference to line A in Table). In order to understand the result and its implication in trading behavior, let us clarify disposition effect and herd behavior in stock market at first.

Shefrin and Statman (1985) identified what they termed the “disposition effect” that investors have a desire to realize gains by selling stocks that have appreciated, but to delay the realization of losses. Odean (1998) tested and demonstrated the disposition effect among investors by analyzing trading records for 10,000 accounts at a large discount brokerage house. Weber and Camerer (1998) designed experiments to test disposition effect validity and explained it by prospect theory (Tversky and Kahneman, 1992). Grinblatt and Keloharju (2001) found evidence that investors are reluctant to realize losses using a unique data set in Finnish stock market. Disposition effect is abnormal behavior in selling stock.

There are innumerable social and economic situations in which we are influenced in our decision making by what others around us are doing (Banerjee, 1992). Herding has been theoretically linked to many economic activities. It is often said to occur when many people take the same action (Graham, 1999). There are many herd behaviors (Hirshleifer and Teoh, 2009), for example, Nofsinger and Sias (1999) studied herding and feedback trading by institutional and individual investors. Lux (1995) formalized herd behavior to explain the emergence of bubbles. Trading conditioning herd is that we are participants with others when expecting return and onlookers with others when expecting loss subject to price volatility and return information in trading market. Herd is abnormal behavior in buying stock.

If, on selling side, the larger the positive mean return is, the more share volume are sold, and the larger the negative mean return is, the less share volume are sold, then, it is disposition effect. If, on buying side, the larger the positive mean return is, the more share volume are bought, and the larger the negative mean return is, the less share volume are bought, then, it is trading conditioning herd behavior.

Therefore, it is true that there are significant disposition effect and trading conditioning herd behavior simultaneously if there is a significant positive correlation between the rate of mean return and that of change in trading conditioning intensity. The magnitude of correlation coefficient indicates intensity in the behavior. Our empirical results show that there is significant positive correlation between the rate of mean return and that of change in trading conditioning intensity in general. Therefore, there are significant disposition effect and trading conditioning herd behavior in stock market in general. It is physiological response and conditioned behavior for return in human beings when people stimulated by price volatility and return information trade in expectancy for return.

### 4.3 Cognitive and Trading Behavior

We subdivide our two year high frequency data into five time intervals based on financial crises in the world and bubble crash in China in 2007. There is a major advantage for this because we can study cognitive and trading behavior change with price volatility, time, and environment.

The correlation coefficients are 0.0729 and 0.1026 right before and just after bubble crashes, respectively. The positive correlation lacks significance. Disposition effect and herd behavior lack significance. It is explained that disagreement on price between bounded rational and irrational traders increases uncertainty in last bubble riding (see line C and D in Table). If we can further demonstrated that there is a relation between bubble crash and the lack of significance in the correlation, we might time and predict bubble crash at a certain extent.

It is more interesting that the correlation coefficient is 0.1963 when SH Index continues dropping in second half time interval after bubble crashes. The positive correlation shows significance. The disposition effect and herd behavior are significance. In comparison with behavior just after bubble crash, we further demonstrate that disposition effect and herd behavior are stronger in a more loss situation (please compare the correlation coefficient in line E with line D in Table).

The correlation coefficient is 0.4766, the largest among 5 subdivided time intervals, and shows significance. It is the most notable disposition effect and herd behavior (see line F in Table). This indicates two facts. First, investors have been conditioned by drop momentum after a year succession of large drop from top 6124.04 points to bottom 1664.04 points in SSE Composite Index. Disposition effect to realize gains shows the most significance in short term. Second, it is a necessary condition for reversal after there is price momentum drop in market that we enhance market confidence and expectancy in return, keep incremental amount of capital into market continuously, and buy a large volume of stock shares that have appreciated and sold by strong disposition effect in short term.

Now, we will focus on a special case that there is a negative correlation, -0.2567, between the rate of mean return and that of change in trading conditioning intensity. It shows significance (see line B in Table).

SSE Composite Index increases almost continuously from 998.23 points at bottom in June, 2005 to 3183.98 points in March, 2007, the beginning of sampling. When investors are conditioned by sustaining positive return, they are reluctant to sell shares with price rising. It is an “anti-disposition effect” (see line B in Table).

We explain variation in the correlation coefficient by that Investors change their expectancy in return and psychological behavior in stock market.

Finally, the correlation coefficient between the rate of mean return and that of change in amount of transaction is always several points (0.03~0.08) higher than that between the rate of mean return and that of change in trading conditioning intensity. Money plays more important role in mean return than psychological behavior and expectancy do in stock market. It demonstrates a dictum in Wall Street: cash is king.

### 4.4 Trading Conditioning and Excessive Trading

We extend three-term contingency operant conditioning to a six-term trading conditioning loop as follows: price volatility and return information (discriminative stimuli) $\rightarrow$cognition, judgment, and decision (decision making) $\rightarrow$trading and no trading (operant) $\rightarrow$expectancy on return (profit and loss) in the future (consequence) $\rightarrow$consequence emotion (new discriminative stimuli) $\rightarrow$ adjustment in expectancy on return (feedback). Please see figures 4 and 7.

Let us go back to reconsider intra-cranial stimulation (ICS) experiment in introduction. Assumed that the rat is rational in bar pressing for foods and irrational for pleasures, we might find a clue why investors behave irrational by excessive trading ignoring wealth loss and maximum utility.

Pierce and Cheney (2004) wrote that money can be exchanged for foods and services, and perhaps is the most important source of economic and conditioned reinforcement. Money and return is positive reinforcement in price volatile trading conditioning in stock market. It has several characteristics as follows: First, it does not loss reinforcement value as quickly as primary reinforcement does because physiological demand is satisfied. Second, profit is a positive reinforcement whereas loss is a negative reinforcement. People trade and buy if they expect price rising. On the other hand, they trade and sell if they expect price dropping. There is trading no matter whether price in expectancy rises or drops. Trading frequency is higher. Third, because we have not yet found time cycle in volume-price probability wave, profit occurs after stock holding in an uncertain time interval. This makes tremendous resistance to extinction in trading conditioning, similar to variable interval (VI) schedules in operant conditioning experiment (Coon, 2007). It can increase trading probability and frequency if people not leave market, for example, loss stopping traders are likely to buy stock again, stimulated by price volatility and positive return (herd behavior). Forth, the trading consequence (profit and loss) that produces feedback to traders could influence their emotion and judgment, and enhance or change their expectancy in return to make trading. Expectancy change also produces high frequency trading. For examples, just buying traders may sell their stock immediately because they find market movement in an opposite direction. Investors who have realized gains by selling stock may become overconfidence and buy stock again. Trading will not stop. In conclusion, we find that trading conditioning produces excessive trading (volume) in stock market.

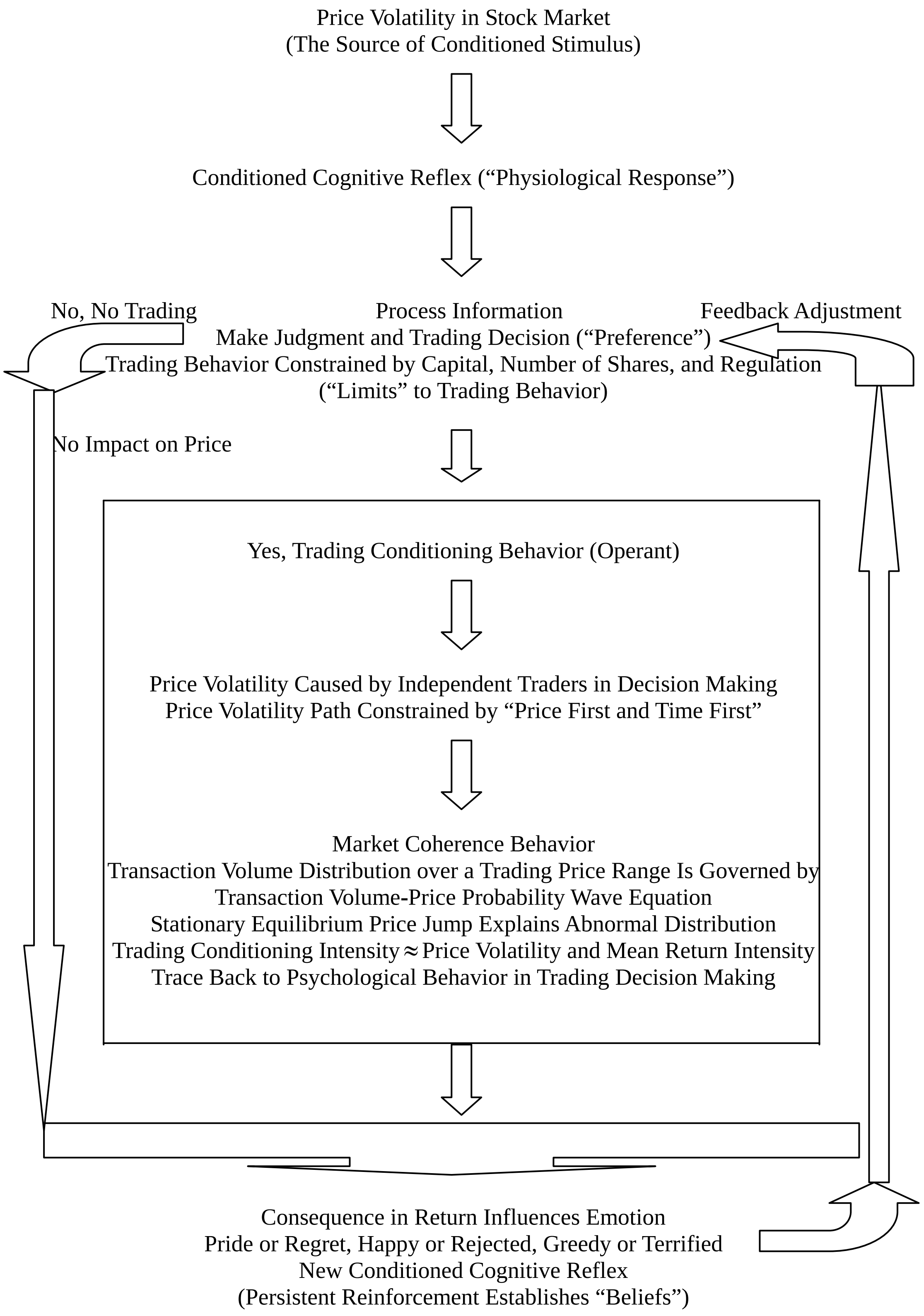

Figure 7: Excessive trading mechanism in a trading conditioning loop

## 4.5 Finance Theory in Interdisciplinary Fields

Shiller (2006) summarized the history of financial theory over the last half century in terms of two distinct revolutions. The first was the revolution by neoclassical finance that began around 1960s, and the second was the revolution by behavioral finance which began in the 1980s. Unlike neoclassical finance that assumed investors' rationality and maximum utility in decision making, behavioral finance attempts to establish a link among neoclassical finance, psychology, and decision making science, study how investors make systematic biases in cognition, judgment, and decision, and explain anomalies in financial market from psychological behavior (Tversky and Kahneman, 1973). Today, we have made great progress on how preferences, beliefs, and limits to arbitrage influence financial market, respectively. Of all non-expected utility (EU) theories, prospect theory (Kahneman and Tversky,1979)(Tversky and Kahneman, 1992) may be the most promising (Barberis and Thaler, 2003).

It is necessary for newly emerging behavioral finance to have better explanations that we need develop a unified hypothesis, methodology, and theory about financial economics while incorporating salient aspect of human behavior so that both rational and irrational behaviors are covered in extreme conditions (Dong, 2009).

Barberis and Thaler (2003) concluded that behavioral models typically capture something about investors' beliefs, or their preference, or the limits to arbitrage, but not all three. In addition, there are obviously competing behavioral explanations for some of the empirical facts.

Shiller (2006) continued his viewpoint that behavioral finance is not wholly different from neoclassical finance. Perhaps the best way to describe the difference is that behavioral finance is more eclectic, more willing to learn from other social sciences and less concerned about elegance of models and more with the evidence that they describe actual human behavior.

We attempt to use econophysics methodology to link physiology, cognitive psychology, neuroeconomics (Camerer et al., 2005), probability and statistics, economic finance, behavioral finance, and econophysics together to study financial market, a multi-interdisciplinary science (see figure 7).

## 4.6 Possible Applications

There are many possible applications in our research. First, we find that trading conditioning produces excessive trading (volume) in stock market. This can help us to explain why and how personality, e.g. overconfidence, sensation seeking, and disagreement, produces excessive trading and high volume from the viewpoints in cognitive psychology, neuroeconomics (Hsu et al., 2005), and medicine. For example, it supports disagreement excessive trading hypothesis (Hong and Stein, 2007) in a certain extent because of positive correlation between the rate of mean return and the rate of change in trading conditioning intensity in general. Second, we use high frequency data to test disposition effect and herd behavior simultaneously by positive correlation between the rate of return and that of change in trading conditioning intensity. It provides a new quantitative analysis and test model for behavioral finance. Third, if we can further demonstrated that there is a relation between bubble crash and the lack of significance in correlation between the rate of mean return and that of change in trading conditioning intensity, we might time and predict bubble crash at a certain extent. In this way, we can improve investment strategy, avoid large risk, and capture good opportunity in investment. Forth, policy makers may use reinforcement to manage irrational behavior and reduce risk in financial market (Xiao and Houser, 2005). Fifth, it is the best laboratory to do research on conditioning in physiology, psychology, and medicine field.

# 5. SUMMARIES AND MAIN CONCLUSIONS

In a real financial market, there are a great many factors that could influence investors' subjective cognition and transaction volume. Therefore, it is a key to develop a theoretical trading conditioning model subject to price volatility and return information how we incorporate market subjective trading behavior into transaction volume-price probability wave distribution function.

Based on classical conditioning and Skinners' three-term contingency operant conditioning, we define price volatile trading conditioning in stock market as participants trade (operant) and expect return in the future (consequence) after they analyze, judge, and have decision making (cognition) in the presence of price volatility and return information (discriminative stimuli). The trading consequence that produces feedback (feedback) to influences their emotion and judgment, adjusts their expectancy in return (emotion), and makes trading.

We develop a theoretical trading conditioning model subject to price volatility and return information in terms of market psychological behavior, based on normative analytical transaction volume-price probability wave distribution function in econophysics. We use transaction volume probability to represent trading conditioning intensity in the model.

In the same way, when we use stationary equilibrium price jump to represent price volatility mean return in any two consecutive trading days, we can use the rate of change in total transaction volume to describe the rate of change in trading conditioning intensity subject to the rate of mean return and its information in the two days. In this way, we can use high frequency data to study correlation between the rate of mean return and that of change in trading conditioning intensity, to analyze psychological behavior change when participants make trading decision in stock market, and to provide a quantitative analysis method and test model for behavioral finance.

In empiric test, we use every trading high frequency data in SSE Huaxia 50ETF from April 2, 2007 to April 10, 2009. We study stationary equilibrium at first in stock market and further evidence the early finding that stationary equilibrium exists widely and transaction volume-price probability wave distribution holds true. Second, we get many good and useful results by studying the correlation between the rate of mean return and that of change in trading conditioning intensity using high frequency data. They are, for examples, (a) we test and explain disposition effect and herd behavior using significant positive correlation. General speaking, there are significant disposition effect and herd behavior simultaneously. We explain them by trading conditioning theory. It is physiological response and return conditioned behavior in human beings when people stimulated by price volatility and return information trade in expectancy for return; (b) we find anti-disposition effect and explain it by sustainable positive return (conditioned reinforcement) rewarded to buying and holding behavior in bull market; (c) it is a necessary condition for reversal after there is price momentum drop in market that we enhance market confidence and expectancy in return, keep incremental amount of capital into market continuously, and buy a large volume of stock shares that have appreciated and sold by strong disposition effect in short term; (d) it lacks significance in spite of positive correlation in two time intervals right before and just after bubble crashes. Probably, it will help us to time bubble crash in stock market; (e) the magnitude of correlation coefficient between the rate of mean return and that of change in the amount of transaction is always several percent points (0.03~0.08) higher than that between the rate of mean return and that of change in trading conditioning intensity. We can draw an inference that money plays more important role in mean return than psychological behavior and expectancy do in stock market. Third, trading conditioning produces excessive trading.

References:

Ait-Sahalia, Yacine (2004): "Disentangling Diffusion from Jump," *Journal of Financial Economics*, 74, 487-528.

Benos, Alexandros V. (1998): "Aggressiveness and Survival of Overconfident Traders," *Journal of Financial Markets*, 1, 353-383

Barber, Brad M., AND Terrance Odean (2000): "Trading Is Hazardous to Your Wealth: The Common Stock Investment Performance," *Journal of Finance*, 55, 773-806.

Barber, Brad M., Terrance Odean, AND Ning Zhu (2009): "Systematic Noise," *Journal of Financial Markets*, 12, 547-569.

Barberis, Nicholas, AND Richard Thaler (2003): "A Survey of Behavioral Finance," *Handbook of the Economics of Finance (Edited by G.M. Constantinides, M. Harris and R. Stulz)*, Elsevier Science B.V., 1051-1121.

Camerer, Colin F., George Loewenstein, AND Drazen Prelec (2005): "Neurosciences: How neuroscience can inform economics?" *Journal of Economic Literature*, XLIII, 9-64.

Coon, Dennis, AND John O. Mitterer (2007): Introduction to Psychology—Gateways to Mind and Behavior (11th Edition), Brooks/Cole Publishing Company.

Cowles, J.T. (1937): "Food-tokens as Incentive for Learning by Chimpanzees," *Comparative Psychology Monographs*, 14 (5, Whole no. 71).

Dong, Zhiyong (2009): Behavioral Finance, Peking University Press, 303-304.

Dragoi, V. (1997): "A Dynamic Theory of Acquisition and Extinction in Operant Learning," Neural Networks, 10, 201-229.

Graham, John R. (1999): "Herding among Investment Newsletters: Theory and Evidence," *Journal of Finance*, 54, 237-268.

Graham, John R., Campbell R. Harvey, AND Hai Huang (2009): "Investor Competence, Trading Frequency, and Home Bias," *Management Science*, 55, 1094-1106.

Greenwood, D.T. (1977): Classical Dynamics (2nd edition), Prentice-Hall, Englewood Cliffs, NJ.

Grinblatt, Mark, AND Matti Keloharju (2001): "What Makes Investors Trade?" *Journal of Finance*, 56, 589-616.

Grinblatt, Mark, AND Matti Keloharju (2009): "Sensation Seeking, Overconfidence, and Trading Activity," *Journal of Finance*, 64, 549-578.

Hirshleifer, David AND Siew Hong Teoh (2009): "Thought and Behavior Contagion in Capital Markets," Handbook of Financial Markets: Dynamics and Evolution (edited by Hens and Schenk-Hoppe), North-Holland/Elsevier.

Hong, Harrison, AND Jeremy C. Stein (2007): "Disagreement and Stock Market," *Journal of Economic Perspectives*, 21, 109-128.

Hsu, Ming, Meghana Bhatt, Ralph Adolphs, Daniel Tranel, AND Colin F. Camerer (2005): "Neural Systems Responding to Degrees of Uncertainty in Human Decision-Making," *Science*, 310 (5754), 1680-1683.

Irons, Jessica G., and William Buskist (2008): "Operant Conditioning," 21st Century Psychology—A Reference Handbook (ed. by Davis and Buskist), 329-339.

Kahneman, Daniel, AND Amos Tversky (1979): "Prospect Theory: An Analysis of Decision under Risk," *Econometrica*, 47, 263-292.

Lee, Charles M.C., AND Bhaskaran Swaminathan (2000): "Price Momentum and Trading Volume," *Journal of Finance*, 55, 2017-2069.

Li, Haitao, Martin T. Wells, AND Cindy L. Yu (2008): "A Bayesian Analysis of Return Dynamics with Levy Jumps," *The Review of Financial Studies*, 21, No.5, 2345-2378.

Lo, Andrew W., AND Jiang Wang (2006): "Trading volume: Implications of an Intertemporal Capital Asset Pricing Model," *Journal of Finance*, 61, 2805-2840.

Lux, Thomas (1995): "Herd Behavior, Bubbles and Crashes," *The Economic Journal*, 105 (July), 881-896.

McCauley J. L. (2000): "The Futility of Utility: How Market Dynamics marginalize Adam Smith," *Physica A*, 285, 506-538.

Nofsinger, John R. AND Richard W. Sias (1999): "Herding and Feedback Trading by Institutional and Individual Investors," *Journal of Finance*, 54 (6), 2263-2295.

Odean, Terrance (1998a): "Volume, Volatility, Price, and Profit When All Traders Are Above Average," *Journal of Finance*, 53, 1887-1934.

Odean, Terrance (1998b): "Are Investors Reluctant to Realize Their Losses?" *Journal of Finance*, 53, 1775-1798.

Odean, Terrance (1999): "Do Investors Trade Too Much?" The American Economic Review, 89, 1279-1298.

Olds, M.E. AND J.L. Fobes (1981): "The Central Basis of Motivation: Intracranial Self-stimulation Studies," *Annual Review of Psychology*, 32, 523-574.

Osborne, M. F. M. (1977): The Stock Market and Finance from a Physicist's Viewpoint, Grossga, Mineapolis.

Pavlov, Ivan (1904): "Physiology of Digestion," *Nobel Lectures—Physiology or Medicine 1901-1921*, Elsevier Publishing Company, Amsterdam, 1967.
(available at http://nobelprize.org/nobel_prizes/medicine/laureates/1904/pavlov-lecture.html )

Pierce, W.D., AND C.D. Cheney (2004): Behavior Analysis and Learning (3rd.), Mahwah, NJ: Erlbaum.

Sewell, Martin (2008): "Behavioral Finance," working paper.

Shefrin, Hersh, AND Meir Statman (1985): "The Disposition to Sell Winners Too Early and Ride Losers to Long: Theory and Evidence," *Journal of Finance*, 40, 777-790.

Shi, Leilei (2006): “Does Security Transaction Volume-price Behavior resemble a Probability Wave?” *Physica A*, 366, 419-436.
Shiller， Robert J. (2006): “Tools for Financial Innovation: Neoclassical versus Behavioral Finance,” *The Financial Review*, 41, 1-8.
Skinner, B.F. (1938): The Behavior of Organisms: An Experimental Analysis, New York: Appleton-Century-Crofts.
Soros, George (1995): The Alchemy of Finance—Reading the Mind of the Market, International Publishing Co..
Statman, Meir, Steven Thorley, AND Keith Vorkink (2006): “Investor Overconfidence and Trading Volume,” *The Review of Financial Studies*, 19, 1531-1565.
Thorndike, E.L. (1913): Educational Psychology (Vol. 2), New York: Teachers College.
Tversky, Amos, AND Daniel Kahneman (1973): “Availability: A Heuristic for Judging Frequency and Probability,” *Cognitive Psychology*, 5, 207-232.
Tversky, Amos, AND Daniel Kahneman (1992): “Advances in Prospect Theory: Cumulative Representation of Uncertainty,” Journal of Risk and Uncertainty,” 5, 297-323.
Weber, Martin, AND Colin F. Camerer (1998): “The disposition effect in securities trading: an experimental analysis,” *Journal of Economic Behavior and Organization*, 33, Issue 2, 167-184.
Xiao, Erte, AND Daniel Houser (2005): “Emotion Expression in Human Punishment Behavior,” Proceedings of the National Academy of Sciences of the United States of America, 102 (20), 7398-7401.

# FOLLOWED IS CHINESE VERSION

# 证券价格波动的交易性条件反射模型

石磊磊*[1]、王毅文[2]、陈定[3]、韩立岩[2]、朴燕、苟成玲[4]

[1]中国科学技术大学近代物理系复杂系统研究组
[2]北京航空航天大学金融系
[3]嘉实基金管理公司
[4]北京航空航天大学物理系

（2010 年 2 月 8 日修改稿）
（欢迎讨论）

摘要

基于解析的成交量价概率波分布函数，用成交量概率描述价格波动的不确定性和强度，我们根据市场群体的心理行为，构造了一个关于价格波动和收益信息的交易性条件反射的理论模型。应用此模型对中国股市高频数据检验，我们主要有以下发现：1）总体来说平均收益率与交易性条件反射强度变化率之间存在着显著的正相关；2）在泡沫破裂前、后两个时期，它们之间的正相关缺乏显著性；3）比较特殊的是上证指数在牛市上升的一段时期内，它们之间存在着显著的负相关。我们的模型和实证能够同时检验股市中的处置效应和羊群行为，并且解释过度交易等市场异象。

*联系作者：在某著名证券公司工作，电子信箱 Shileilei8@yahoo.com.cn 或leilei.shi@hotmail.com
作者感谢与王怀玉、王亚平、周春生、刘劲和刘发民等非常有意义的讨论。

A SECURITY PRICE VOLATILE TRADING CONDITIONING MODEL

Leilei Shi*[1], Yiwen Wang[2], Ding Chen[3], Liyan Han[2], Yan Piao, and Chengling Gou[4]

[1]Complex System Research Group, Department of Modern Physics
University of Science and Technology of China
[2]Department of Finance, Beijing University of Aeronautics and Astronautics
[3]Harvest Fund Management Co. Ltd.
[4]Department of Physics, Beijing University of Aeronautics and Astronautics

Abstract

We develop a theoretical trading conditioning model subject to price volatility and return information in terms of market psychological behavior, based on analytical transaction volume-price probability wave distributions in which we use transaction volume probability to describe price volatility uncertainty and intensity. Applying the model to high frequent data test in China stock market, we have main findings as follows: 1) there is, in general, significant positive correlation between the rate of mean return and that of change in trading conditioning intensity; 2) it lacks significance in spite of positive correlation in two time intervals right before and just after bubble crashes; and 3) it shows, particularly, significant negative correlation in a time interval when SSE Composite Index is rising during bull market. Our Model and findings can test both disposition effect and herd behavior simultaneously, and explain excessive trading (volume) and other anomalies in stock market.

*Corresponding author. E-mail address: Shileilei8@yahoo.com.cn or leilei.shi@hotmail.com, working in an international securities company in China.

# 一、引言

长期以来，人们都十分关注价格波动行为的研究，而忽略了对交易量的认识。近 10 年来这种现象有所改变，人们开始逐步重视和研究在金融交易市场中交易量所包含的内容和信息。在新古典金融理论的框架内，Lo 和 Wang（2006）试图通过建立价格和交易量与基本面之间某种联系来强调应该把量价作为一个整体对资本市场进行分析。在新兴发展的行为金融学范畴内，我们开始将交易量与投资者的情绪、信念和偏好联系在一起。Sewell（2008）认为：行为金融学是研究心理学对金融市场参与者行为的影响并且由此产生的市场效应，它有助于解释市场为什么以及怎样无效的。Lee 和 Swaminathan（2000）发现以前的成交量在趋势投资策略和价值投资策略之间提供了重要的连接，它有益于调和价格在中期反应不足而在长期反应过度的效应；Benos（1998）和 Odean（1998）认为过于自信（overconfidence）导致市场的过度交易（excessive trading）；Odean（1999）例举三个理由说明过于自信的投资者产生过度的交易量：职业选择偏差、生存能力偏差以及对预期收益的不现实理念，并通过对客户交易收益和交易成本的研究来检验过于自信导致过度交易的假说；Barber 和 Odean（2000）进一步的实证发现活跃的交易行为会降低收益率，这种非理性行为只能用过于自信来解释；Graham 等（2009）则从那些自认为比别人更具竞争力的群体交易更加频繁来解释过于自信导致过度交易假说。Statman 等（2006）用过于自信模型对交易量的预测性进行检验，发现个股的换手率与滞后回报率以及与滞后市场回报率之间持续数月的关系分别显示了处置效应和过于自信。Barber 等（2009）用个人投资者系统地买入近期表现好的股票而不卖亏损的股票（使得净交易量非常大）的心理偏差来解释个体对价格表现的交易行为存在着高度持续的相关性。Grinblatt 和 Keloharju（2001）通过实证发现过去的回报率、参照价格效应、避税卖出和持有期间盈亏的大小都是影响交易的因素，而过于自信和寻求兴奋感（sensation seeking）的人交易都更加频繁和活跃（Grinblatt 和 Keloharju，2009）。Hong 和 Stein（2007）认为交易量似乎是反应投资者观点的指标，价格偏离基本价值越高，交易者对价格的分歧就越大，交易量就表现出异常地大，由此，提出了一个可以将股价和交易量统筹考虑的意见分歧交易量模型（disagreement model）。

无论是过于自信交易量假说，还是寻求兴奋感或意见分歧交易量假说都认为交易量的大小反应了投资者情绪、信念和偏好的强弱程度。

1904 年诺贝尔生理与医学奖获得者、俄国生理学家巴甫洛夫（Pavlov，1904）在研究动物生理反应和反射时用狗的唾液量来表示反射的强度，提出了经典性条件反射。经典性条件反射是当条件反射刺激出现时预期非条件反射刺激物的一种生理反应。

Thorndike（1913）最早研究了操作性条件反射。Skinner（1938）为了研究大鼠操作性条件反射专门设计了一个装置，“斯金纳箱”（Skinner box）。实验发现：经过几次操作获取食物（强化物）之后，大鼠便形成了稳定的频率按压杠杆获取食物的行为模式。现在，心理学家把操作性强化物定义为“伴随一个反应发生并能增加这个反应再次发生的可能性事件”。

与经典条件反射一样，操作性学习也是建立在信息和期望基础之上的。操作性条件反射是在特定时间通过特定操作预期得到特定结果（强化物）的行为反应。Pierce 和 Cheney（2004）把操作性条件反射归纳成“由特定刺激营造的时机从事操作，进而产生某种后果（强化物）”的行为反应。或者说在特定的刺激出现时发生操作性条件反射行为，并伴随着特定的结果 (Irons and Buskist, 2008)。Skinner 将这种关系称之为三项相互联系、可能发生的事件（three-term contingency）（Dragoi，1997）。在操作性条件反射中，强化物可被用于改变反应频率：强化物出现的不确定性（例如不确定反应次数后出现强化物的“可变比率模式”和不确定时间间隔后出现强化物的“不定时间间隔模式”）使得条件反射的频率很高、消退的抑制力很强（库恩，2007）。

颅内刺激（intracranial stimulation, ICS）能够直接激活大脑中的快乐中枢，有着最强、最特殊的强化作用（Olds 和 Fobes，1981）。实验发现许多大鼠为获得愉快感每小时能按压杠杆几千次，远远大于为获取食物按压的频率（相差几个数量级），全然不顾对食物、水和性的需求！

库恩（Coon，2007）将操作性强化物分成三类：一级强化物、二级强化物和反馈。一级强化物是自然形成的和非习得性的，具有生理基础，能产生舒适感和消除不适感，或能满

足即时的生理需求。金钱、赞扬、赞同、情感以及其他的奖赏都可以成为强化物，即习得的二级强化物。一些二级强化物具有“代币物”的功能，可用以交换一级强化物，因此就有了更直接的价值。例如，钱币本身没有太大的价值，既不能充饥也不能解渴，但是，我们可以用它换取商品和服务，也许是最重要的经济强化物（Pierce 和 Cheney，2004）。研究者曾通过实验教黑猩猩为获得代币物而工作（Cowles，1937）。反馈是指人们对操作反应结果信息的接收和处理之后再影响操作的一个过程。

索罗斯（Soros，1995）最初从抽象的哲学思考出发研究反射理论，随后逐步察觉到它与股票价格行为的相关性。他认为股票市场可以成为研究反射现象的最佳起点。索罗斯定性描述的反射是指人们参与交易导致价格波动，通过重新对交易价格的认知、思考和判断再参与交易的反馈过程，属库恩（Coon，2007）分类中的第三类操作性强化物。

在研究每日证券成交量在价格波动区间分布整体行为时，Shi（2006）用规范的经济物理方法推导出一个解析的成交量价波动方程并且得到两组解析的成交量在价格波动区间的分布函数，用成交量概率描述价格波动的不确定性和强度。基于此，我们根据市场群体的心理行为用成交量概率描述价格波动的交易性条件反射强度，构造了一个关于价格波动和收益信息的交易性条件反射的理论模型。通过该模型研究平均收益率与交易性条件反射强度变化率的相关性，我们同时检验了股市中的处置效应和羊群行为，由此追溯和分析市场群体在做交易决策时因价格波动和信息环境不同而发生的心理行为的变化，解释过度交易（量）等市场异象，试图为行为金融提供一个新的分析和检验模型。

尽管在过去有许多关于回报与交易量的研究，也有一些交易量行为的研究，但是据笔者所知还没有回报率与成交量（概率）变化率关系的研究，没有将经济物理学中严谨规范的解析量价概率波分布与生理学和心理学中解释性的认知条件反射行为成功地结合在一起进行研究——绝大多数试图同时实现规范和解释性目标的非期望效用理论模型的困难之处是它们最终在这两个方面都做的不好（Barberis 和 Thaler，2003）。与成交量不同，成交量变化率或成交量概率变化可以是正的、也可以是负的、或者是零。与过去采用事件研究法和心理学实验方法不同，我们用高频数据同时检验处置效应和羊群行为，并且解释过度交易行为等市场异象。

以下文章的结构是：第二部分是成交量价概率波分布函数简介并且找到了一种关于价格波动和收益信息的交易性条件反射强度的计量方法；第三部分是以华夏上证 50ETF 每笔交易的高频数据为例，我们首先研究定态均衡、确定定态均衡价格、进一步佐证了在股票市场中普遍存在着定态均衡状态（Shi，2006）；之后，我们实证检验了前后交易日之间平均收益率、交易性条件反射强度变化率和成交金额变化率三者之间的相关性；第四部分是实证分析和讨论，这包括 1）市场群体相互作用和干涉结果出现的定态均衡；2）总体来说，市场同时具有显著的处置效应和羊群行为；3）人们的认知和交易行为随价格波动和环境不同发生的变化；4）用交易性条件反射理论解释过度交易；5）应用前景等等；最后是总结和主要结论。

## 二、成交量价概率波分布函数及交易性条件反射强度

### 2.1　成交量价概率波分布函数

在用经济物理的方法研究股票成交量价波动行为时，Shi（2006）观察到在股票交易市场中普遍存在着定态均衡现象。股票每日在价格波动区间的成交量（累计交易量）分布在价格加权平均值附近出现峰化，这种现象与股票价格高低、价格波动路径以及涨跌没有关系[1]。我们将成交量峰值所对应的价格定义为该日的定态均衡价格。另外，价格总是围绕定态均衡价格上下波动的同时，定态均衡价格表现出跳跃（jump）的变化[2]。

---

[1] Shi（2006）对 618 个成交量价分布进行拟合和显著性检验。这些样本价格不同、价格波动的路径和方向也不同，但是都表现出在成交量加权价格平均值附近出现峰化现象。

[2] 与Poisson-diffusion（Ait-Sahalia，2004）和Levy（Li，Wells & Yu, 2008）等跳跃不同，定态均衡价格有清晰的跳跃机制。例如，当有大量资金进入市场买入股票，供需关系突然发生大的变化、打破平衡时，供需关系就会通过价格调节作用寻找新的价格平衡点，定态均衡价格随之出现不连续的跳跃；通过成交量在价

影响股票价格波动的因素有许多，例如，公司经营状况、宏观经济因素、政策因素、心理行为因素等等。它们对价格波动的影响有大有小。但是，各种因素只有通过交易传导才能造成价格的波动，是间接的作用。因此，Shi（2006）在前期建立价格围绕定态均衡价格波动模型时暂时没有考虑这些因素。

没有交易就没有价格的波动，价格波动的同时一定存在交易，但是有交易不一定就有价格的波动，交易是价格波动的必要条件。除此之外，成交金额约束着价格的波动和成交量的大小。在某一价格处的成交量 $v$ 等于在此价格处的累计交易量。在研究定态均衡条件下成交金额是如何定量约束量价行为以及成交量是怎样在价格波动区间分布时，Shi（2006）提出了成交金额流动约束量、成交金额流动分布量和成交金额流动波动量（定态均衡价格回归量）之间相互关系假说，即成交金额约束关系假说：

$$-E + p\frac{v_t^2}{V} + W(p) = 0 \tag{1}$$

其中 $p$ 是价格，其量纲是[货币单位][股]$^{-1}$；$v_t = v/t$ 是在时间间隔 $t$ 内，在价格 $p$ 处的成交量（累计交易量）、动量（momentum）和趋势，其量纲是[股][时间]$^{-1}$；$v/V$ 是在价格 $p$ 处的成交量概率，它等于在价格 $p$ 处的成交量 $v$ 除以总成交量 $V$；$E = p \cdot v/t^2$ 是在时间间隔 $t$ 内，在价格 $p$ 处的成交金额（累计交易金额），即成交金额流动约束量，其量纲是[货币单位][时间]$^{-2}$；$v_{tt} = v/t^2$ 是在时间间隔 $t$ 内，在价格 $p$ 处的冲量和偏离定态均衡价格的趋势力，其量纲是[股][时间]$^{-2}$；$p \cdot v_t^2/V$ 是在时间间隔 $t$ 内，在价格 $p$ 处成交金额流动分布量，其量纲是[货币单位][时间]$^{-2}$；另外，

$$W(p) = A(p - p_0) \approx A\left(p - \overline{p}\right) \tag{2}$$

是在时间间隔 $t$ 内，在价格 $p$ 处成交金额流动回归量，其量纲也是[货币单位][时间]$^{-2}$；$p_0$ 代表在时间间隔 $t$ 内的定态均衡价格，$\left|\overline{p}\right|$ 是价格加权平均值，$A = (1 - v/V)v_{tt}$ 代表定态均衡回归力和供需回归力的大小，其量纲是[股][时间]$^{-2}$。

根据股票交易“价格优先，时间优先”规则，我们有交易价格最小波动定律：股票交易价格波动的路径要求成交金额约束假说方程（1）的泛函

$$F(p,\psi) = (W - E)\psi^*\psi + \frac{B^2}{V}p\left(\frac{d\psi^*}{dp}\right)\left(\frac{d\psi}{dp}\right) \tag{3}$$

在各种可能的价格变化中，选择对当前价格变化最小，即对描述价格波动的函数 $\psi(p)$ 取极值；其数学表达式是

$$\delta\int F(p,\psi)dp = 0。 \tag{4}$$

由方程（1）和方程（4），我们得到量价波动方程

$$\frac{B^2}{V}\left(p\frac{d^2\psi}{dp^2} + \frac{d\psi}{dp}\right) + [E - W(p)]\psi = 0 \tag{5}$$

将方程（2）代入方程（5），并利用自然边界条件

$$\psi(0) = 0、\psi(p_0) < \infty、及 \psi(+\infty) \to 0，$$

我们得到两组解析解；其一，当偏离定态均衡价格的趋势力 $v_{tt}$ 与供需平衡回归力 $A$ 之和在价格波动区间等于本征值常数、发生相互干涉行为时，我们得到

$$\psi_m(p) = C_m J_0[\omega_m(p - p_0)], \qquad (m = 0,1,2,\cdots), \tag{6}$$

其中

$$\omega_m^2 = v_{tt} - A = \frac{v}{V}v_{tt}, \qquad (\omega_m > 0) \qquad (m = 0,1,2,\cdots) \tag{7}$$

---

格波动区间的分布，我们能够观察和测量其跳跃行为（Shi，2006）。

$\omega_m$ 是大于零的本征值常数，$J_0[\omega_m(p-p_0)]$ 是带本征值的零阶贝塞尔（Bessel）函数，$C_m$ 是一无量纲的常数；其二，当供需平衡回归力在价格波动区间等于本征值常数时，我们得到另外一组多阶本征值函数

$$\psi_m(p)=C_m e^{-\sqrt{A_m}|p-p_0|}\cdot F\left(-n,1,2\sqrt{A_m}|p-p_0|\right),\quad (n,m=0,1,2,\cdots) \qquad (8)$$

其中 $$\sqrt{A_m}=\frac{E_m}{1+2n}=\text{常数}>0,\qquad (n,m=0,1,2,\cdots) \qquad (9)$$

$F\left(-n,1,2\sqrt{A_m}|p-p_0|\right)$ 是一个 $n$ 阶的合流超几何函数或第一类 Kummer 函数。

波函数（6）和（8）的含义是它的绝对值 $|\psi_m(p)|$ 代表在价格 $p$ 处的成交量概率（参见图 1 和图 2，其中横坐标可以代表价格，纵坐标代表成交量分布，原点是定态均衡价格），$C_m$ 是归一化常数。

图 1 和图 2 的分布函数有几点不同：第一，图 1 的分布是随着价格不断偏离定态均衡价格表现出波浪式衰减的，而图 2 除了零阶是指数分布之外，其分布是一种波浪式均匀的，阶数越大，均匀性越显著；第二，图 1 的分布是开放的，能够描述股市中突发的、远离均衡价格的脉冲式交易行为，例如，利益输入或输出交易行为，而图 2 的分布是封闭的，能够描述随机均匀分布行为；第三，除了零阶分布，图 2 分布的本征值常数一般比图 1 的大一、二个数量级。但是，图 1 和图 2 分布的共同点是它们都能够很好地描述指数分布。

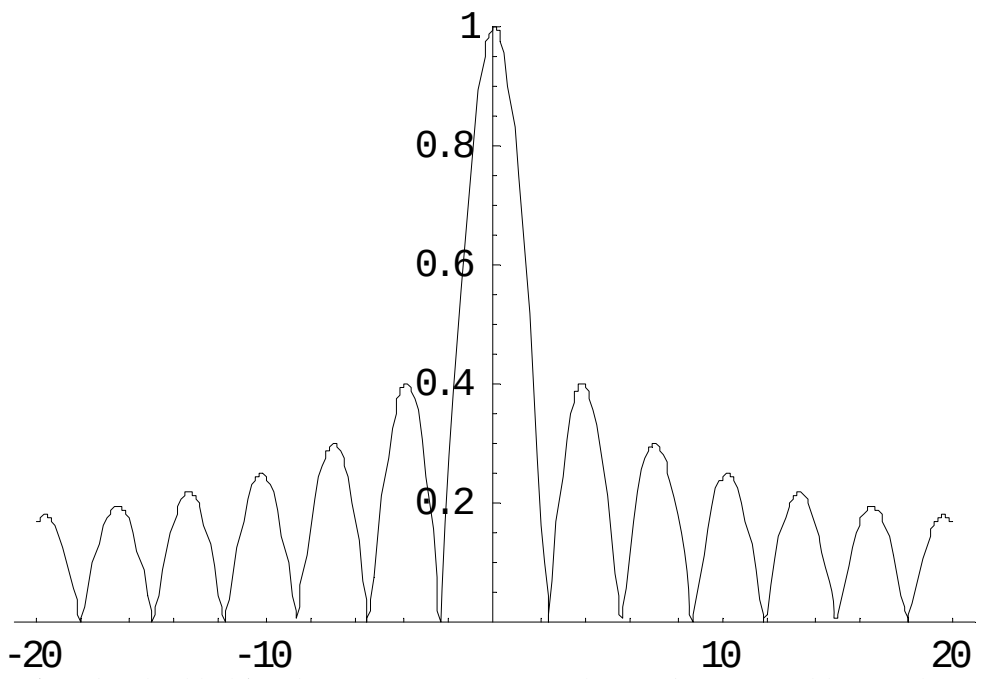

图 1：成交量概率在价格波动区间的零阶贝塞尔函数的绝对值分布函数

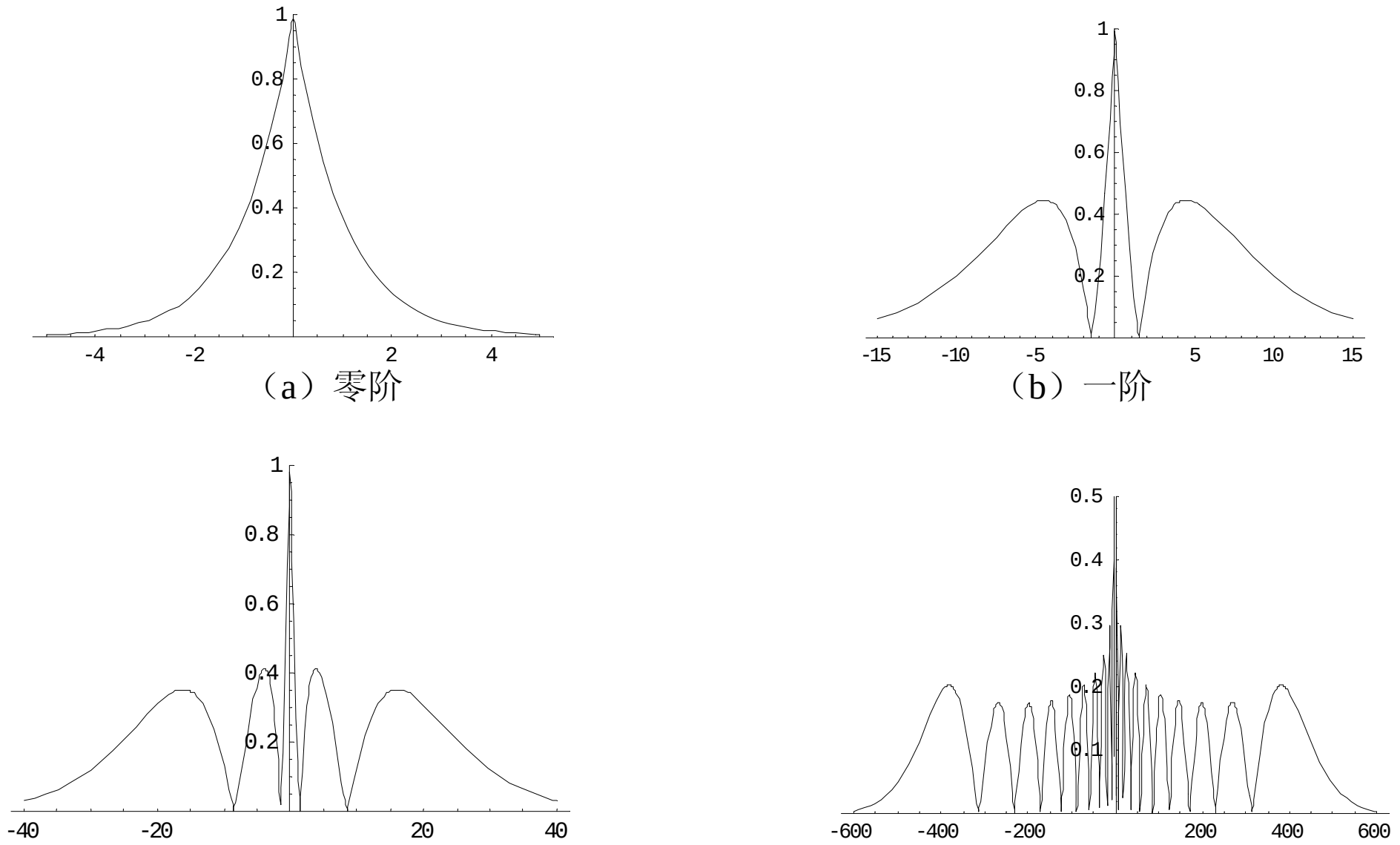
（a）零阶　（b）一阶　（c）二阶　（d）十阶

图 2：成交量概率在价格波动区间的多阶函数的绝对值分布函数

## 2.2　本征值和概率波

本征值又称为特征值。在描述事件分布的函数中，本征值与参数是不同的。在绝大多数情况下，当我们观察到某种事件的分布现象时，由于不清楚导致事件分布的机制，往往用一个带有参数的分布函数描述这种统计行为。在这种近似的描述中，我们并不知道该参数所包含的内容和实际意义。本征值是可观察和可测量的量，明确了产生分布的机制，因此能够对其正确与否进行定量检验。例如，在成交量价概率波分布函数中，本征值包含两种意义：在第一组解中，本征值描述的是偏离定态均衡价格的趋势力 $v_{tt}$ 与供需平衡回归力 $A$ 之和在价格波动区间等于本征值常数、发生相互干涉行为的结果，用方程（7）计量；在第二组解中，本征值描述的是供需平衡回归力在价格波动区间等于本征值常数、出现指数和随机均匀分布的现象，用方程（9）计量。

所谓概率波是用数量分布的概率描述波动的强弱程度，例如在成交量价概率波行为中，我们用成交量分布和概率而不是价格波动的幅度描述价格波动的强度。现在，让我们来比较概率波和经典波的不同点和相同点（参见图 3）。首先，概率波的横坐标是价格或距离，纵坐标是数量概率，而经典波的横坐标是时间，纵坐标是振幅；第二，概率波是用数量分布概率的大小描述波动的强弱和变化，而经典波是用振幅的大小描述波动的强弱和变化；第三，概率波的强度大于或等于零，而经典波的强度可以是负的；第四，相对于平衡位置，概率波波动的强度随着偏离平衡位置呈现出波浪式的变化（除了零阶分布函数），波动的幅度越大，不等于波动的强度也越大，而在经典波中波动的幅度越大，波动的强度就越大；第五，概率波描述群体之间的相互作用，不能独立存在，而经典波可以独立存在；第六，目前我们还没有发现概率波在时间上存在着严格的周期性，即预测重复的时间存在着不确定性，而经典波有传播速度，可以计量，因此存在着时间周期。

概率波和经典波的共性就是都会出现干涉现象，都有周而复始的重复变化。

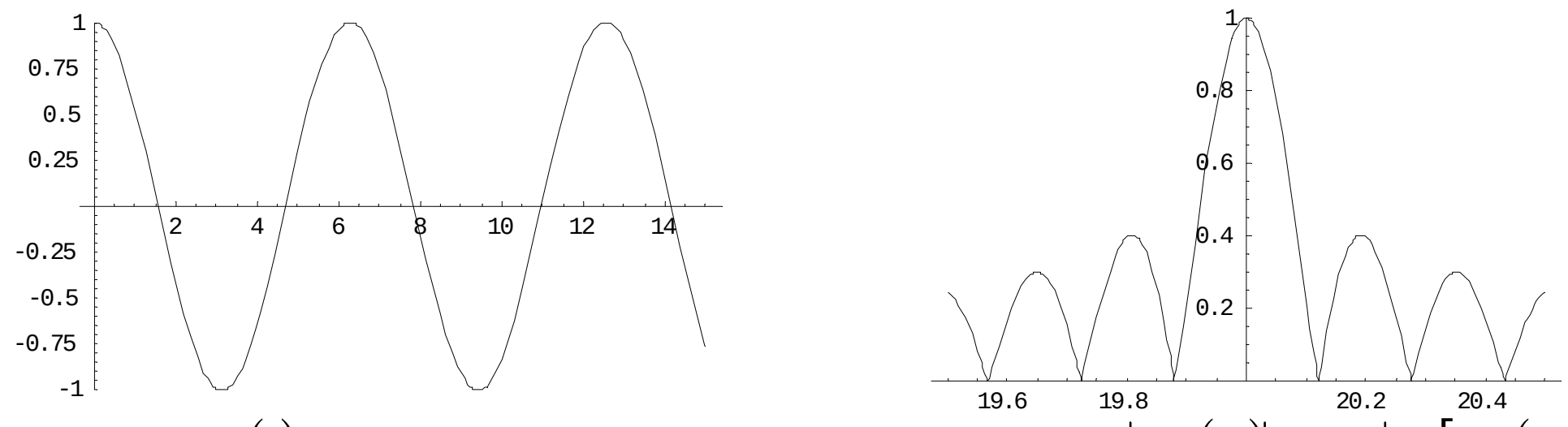

（a）经典波 $y(t)=y_0\cos(\omega t)$　　（b）概率波 $|\psi_m(p)|=C_m|J_0[\omega_m(p-p_0)]|$

图 3：经典波与概率波的比较

## 2.3 交易性条件反射的引入

我们首先考虑没有人们主观变化、没有影响买卖交易行为的刚性交易体系。当交易价格波动时，在任意相同的时间间隔内总成交量是相同的。如果有增量扰动资金进入市场买入股票、打破供需平衡、导致定态均衡价格跳跃，那么增量资金与定态均衡价格跳跃幅度之间存在着一一对应的关系（因为跳跃前后定态均衡的总成交量是相等的）。

在真实的金融交易市场中，有许许多多的因素都会影响人们的主观认知、供需平衡和成交量的变化。如何将市场群体的主观交易行为和成交量变化纳入到解析的成交量价概率波分布函数中，将是认识和解决成交量、价格波动与增量扰动资金之间相互关系的关键，也是建立关于价格波动和收益信息的交易性条件反射理论模型的关键。

人类对衣、食、住、行和服务的生理需求是客观的，对它们的生理反应是一种非条件反射。印币既不能吃，也无法解渴，其本身几乎是没有任何价值。在商品经济中，当我们将货币、收入和收益（率）通过交换与人类赖以生存的物质必需品和服务紧密结合起来时，我们就已经被它们条件化，就会对货币、收入和收益（率）形成同样的生理反应，即经典性条件反射。Pierce 和 Cheney（2004）认为在现代商品经济中钱可用于交换商品和服务，也许是最重要的经济强化物。

在股票交易市场，当人们看到价格波动和收益（盈利或亏损）时产生的对物质必需品和服务需求的生理刺激就是一种经典性条件反射，而根据价格波动和收益信息进行分析、处理、

判断和决策从事买卖交易行为，并预期未来收益（盈利或亏损）的反应是一种操作性条件反射。

交易性条件反射的强化物是钱和收益率，它有以下几个特点。第一，它不会像一级强化物那样因生理需求已经得到满足而失去强化作用；第二，收益对应与正强化物，亏损对应与负强化物。预期股票价格上涨会促使人们买入持有，预期股票价格下跌会促使人们卖出观望。无论是预期价格上涨还是下跌都会出现交易性条件反射，进行交易。第三，由于量价概率波行为没有确定的时间周期，类似与操作性条件反射中部分强化的不定时间间隔模式（库恩，2007），持有股票时间有长有短之后才出现收益，这使得交易性条件反射消退的抑制力很强；第四，损益结果（盈利、亏损或持平）反馈给交易者会影响人们的情绪（emotion）和判断能力，会改变人们对未来收益的预期而导致交易。例如，投资者止损出来后很容易因为看到价格上涨导致收益预期变化再次买入股票，而刚买入股票之后也会因为收益预期变化或已经获得微利又很快卖掉。因此，在交易性条件反射的认知过程中，除了预期收益和亏损外，还包括反馈导致的预期变化。

根据 Skinner 关于三项相互联系、可能发生的事件（three-term contingency）和操作性条件反射的定义（Irons and Buskist，2008），我们定义在股票市场中的交易性条件反射是一种受价格波动和收益信息（discriminative stimuli）影响，通过分析、处理、判断和决策（cognition）进行交易（operant），并预期未来收益（盈利、亏损或持平）（consequence）的一种操作性条件反射；其结果（收益、亏损或持平）反馈（feedback）给交易者会影响他们的情绪（emotion）和判断能力，调整他们的收益预期并进行交易活动（参见图 4）。

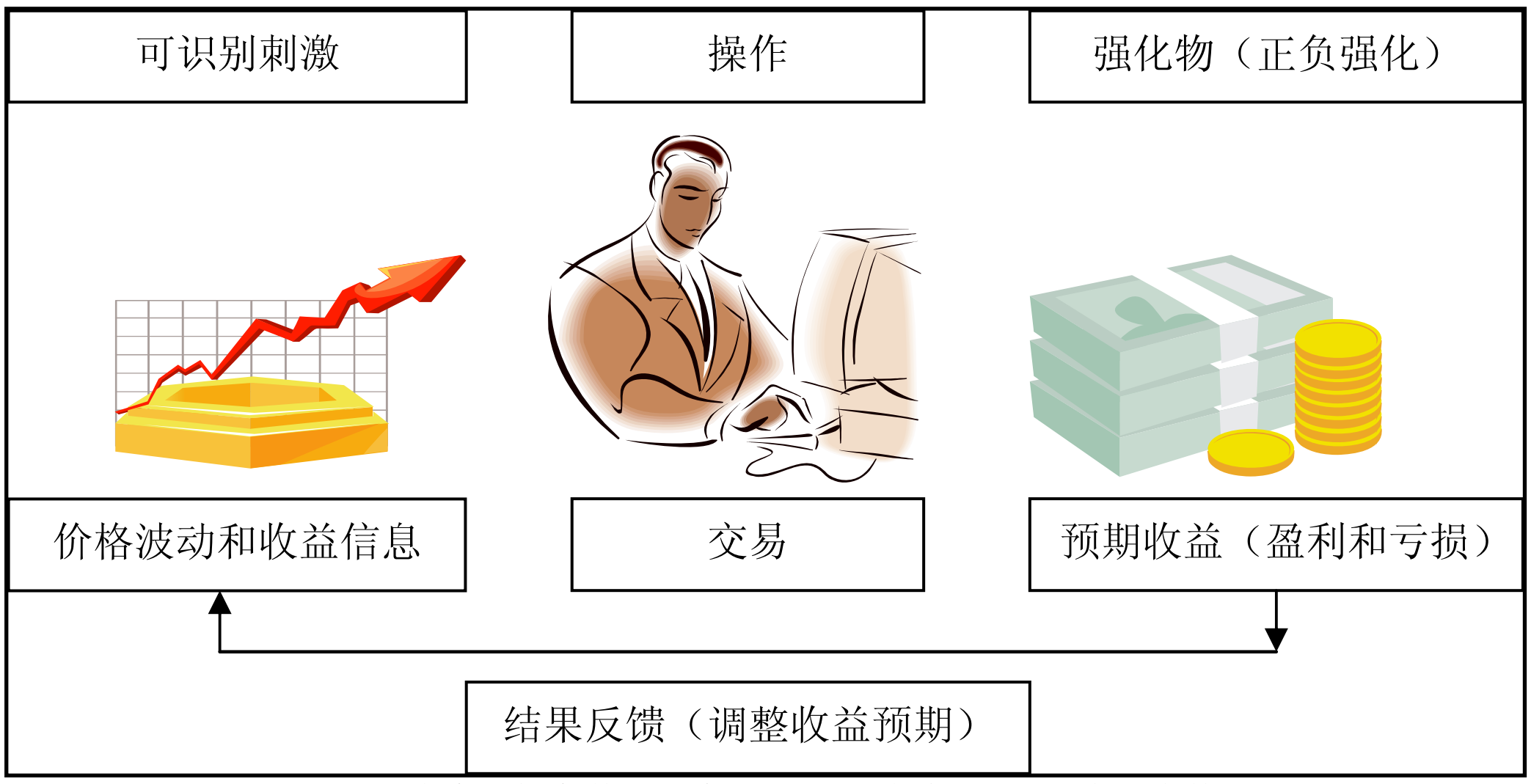

图 4：在交易性条件反射中三项相互联系、可能发生的事件

在股票交易体系中，独立变量有两个，价格和成交量，成交金额是它们的约束量。根据经典力学原理（周衍柏，1985），我们知道交易体系的独立自由度的数目等于 1（独立变量数 2 减去约束方程数 1）。经济物理学先驱 Osborne（1977）发现价格作为供需量的被解释变量（因变量）在实证中是不存在的，McCauley（2000）从数学上证明了 Osborne 的这一发现。因此，Shi（2006）选择价格是股票交易体系中的自变量，成交量是因变量。

根据刚刚建立的交易性条件反射行为，证券的价格波动和收益信息会产生市场群体对该证券未来收益的预期而进行交易。这种主观认知的交易行为和成交量是同步变化的。我们用成交量分布和概率随价格波动的变化（注：不是成交金额的变化，因为成交金额是约束量）表示市场群体受价格波动和收益信息影响，导致交易行为、供需关系和主观认知的变化。

在成交量价概率波分布函数中，我们用成交量概率描述价格波动的强度（不是价格波动的幅度）。在某一价格处，成交量概率越大，波动的强度就越大（注：与经典波不同，在概率波中波动的强度越大不等于波动的幅度就越大，反之亦然），波动的频率就越高[1]，交易频

[1]根据概率统计学的大数定律，当总成交量远远大于单笔成交量时，成交量的概率近似等于频率。因为我们

率也就越高，对未来价格波动、收益和亏损的预期就越高（成交量越大，买量越大，表明市场对未来价格上涨的心理预期越强；成交量越大，卖量越大，表明市场对未来价格下跌的心理预期越强），交易性条件反射的强度也就越大，反之亦然。因此，价格波动的强度近似等于关于价格波动和收益信息的交易性条件反射的强度。我们可以用成交量概率近似地描述交易性条件反射的强度和频率（参见图 5）。

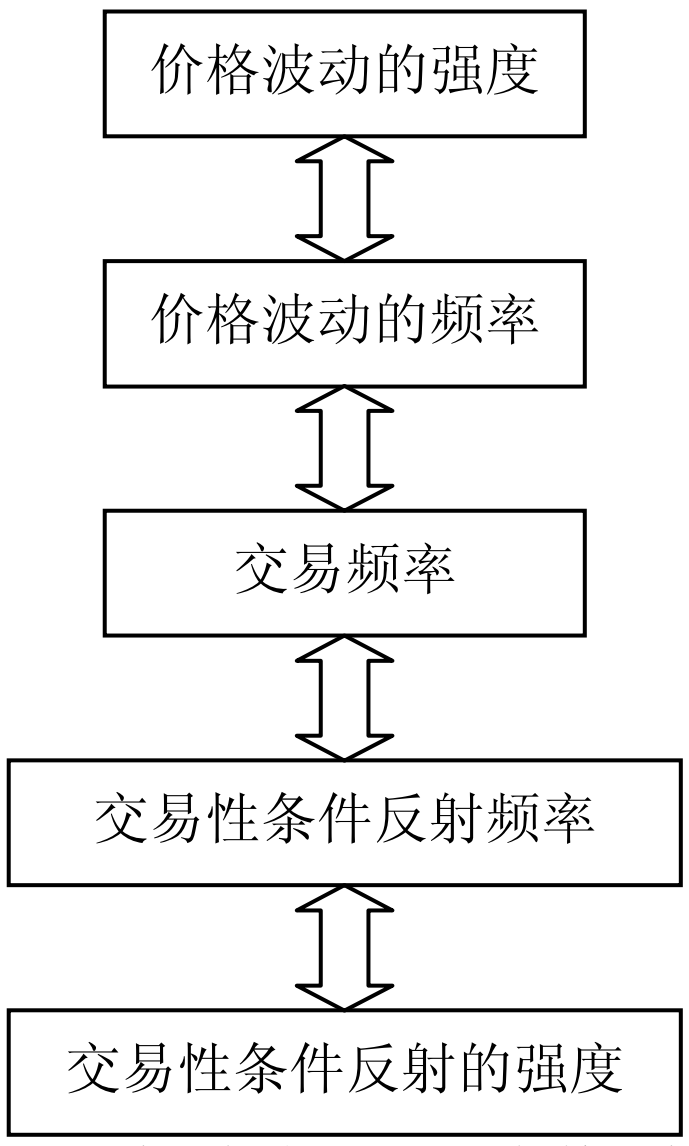

图 5：价格波动的强度（用成交量概率表示）近似等于交易性条件反射的强度

现在，我们用成交量概率表示交易性条件反射的强度。同样，当我们用定态均衡价格跳跃的幅度率表示价格波动的平均收益率时，我们也能够用定态均衡价格跳跃前后成交量的变化率来描述市场群体对该平均收益率的交易性条件反射强度的变化率。显然，交易性条件反射强度的变化率可以是正的、负的或是零。这样，我们就能够通过使用高频数据定量分析平均收益率与交易性条件反射强度变化率之间的相关性，由此研究他们在交易决策时的生理反应和心理行为。

## 三、数据分析及相关性检验

我们以华夏上证 50ETF（510050）每笔交易的高频数据为例，用零阶贝塞尔绝对值分布回归模型对每个交易日的成交量在价格波动区间的分布进行拟合并检验其显著性。对于通过检验的，我们能够从拟合数据中直接得到该交易日的定态均衡价格。对其他交易日，我们用价格加权平均值近似表示该日的定态均衡价格。由此，我们能够计算出前后交易日之间定态均衡价格跳跃的幅度（平均收益率），研究平均收益率、交易性条件反射强度变化率以及成交金额变化率之间的相关性。

### 3.1 数据

我们采用嘉实基金管理公司提供的嘉实高频 HF2 数据库中每笔交易的高频数据。样本区间为 2007 年 4 月 2 日至 2009 年 4 月 10 日。近 740 天，共 495 个交易日，即 495 个量价分布。

我们分两步对原始数据进行了预处理。首先，在原始数据中价格保留了小数点后的三位

---

在下一部分使用样本期间的日平均成交量很大，为 3.6 亿股，所以在某一价格处成交量概率可近似等于交易频率。这样，在某一价格的成交量越大，其概率就越大，频率就越大。对于一笔大单提升的交易量，即使诸多小单对应同样的交易量，虽然两者对应的信息可能不一样，但是成交量概率依然可以近似地描述交易的频率，由此出现的量价分布"异常"现象说明市场资金优势的扰动很容易破坏定态均衡、操纵价格，我们会进一步研究。

数，我们采取四舍五入的原则保留到两位数，同时将相应的成交量相加，得到新的量价数据。接着我们将每一价格处的成交量（累计交易量）除以当天总成交量得到在该价格处的成交量概率。这样，我们就得到了每个交易日的成交量概率在价格波动区间的实际分布。

3.2 成交量价分布拟合、总体显著性检验以及定态均衡价格确定

我们已知在定态均衡状态下，量价分布的理论函数是

$$|\psi_m(p)| = C_m |J_0[\omega_m(p - p_0)]|, \qquad (m = 0,1,2\cdots) \qquad (10)$$

其中 $C_m$、$\omega_m$ 和 $p_0$ 分别是归一化常数、本征值常数和定态均衡价格。它们在单变量非线性回归模型

$$|\psi_m(p_i)| = C_m |J_0[\omega_m(p_i - p_0)]| + \varepsilon_i \qquad (i = 1,2,3\cdots,n) \qquad (11)$$

中是三个待定常系数，$n$ 是样本在交易价格区间内的价格数，$\varepsilon_i$ 是服从 $N(0,\sigma^2)$ 分布的随机误差项，$|\psi_m(p_i)|$ 是在价格波动区间内任一处的成交量概率观察值，$C_m|J_0[\omega_m(p_i - p_0)]|$ 是理论值。我们采用 Levenberg-Marquardt 非线性最小二乘法对每一个量价分布样本进行拟合，确定其中的三个待定常系数并且得到成交量概率分布的理论结果。在拟合分析中，我们使用 Origin6.0 软件（参见图 6（a））。

对总体显著性检验，我们采用 $F$ 检验的方法。可决系数(coefficient of determination)

$$R^2 = \frac{ESS}{TSS} = \frac{TSS - RSS}{TSS}, \qquad (12)$$

其中 $ESS = \sum_{i=1}^{n}(\hat{Y}_i - \bar{Y})^2$ 是回归平方和(explained sum of squares)，$RSS = \sum_{i=1}^{n}(Y_i - \hat{Y}_i)^2$ 是残差平方和(residual sum of squares)，$TSS = \sum_{i=1}^{n}(Y_i - \bar{Y})^2$ 是总离差平方和(total sum of squares)。统计量

$$F = \frac{ESS/k}{RSS/(n-k-1)} \qquad (13)$$

中，$n$ 和 $k$ 分别指样本个数和解释变量个数（独立自由度数）。设定显著性水平 $\alpha = 0.05$，当 $F > F_{0.05}$ 或

$$R^2 > R_{crit}^2 = \frac{k \cdot F_{0.05}}{k \cdot F_{0.05} + (n-k-1)}, \qquad (14)$$

即 $R^2$ 大于临界值 $R_{crit}^2$ 时，则回归模型（11）在 95%的显著性水平下总体显著成立（在这里 $k = 1$）。

检验结果显示自 2007 年 4 月 2 日至 2009 年 4 月 10 日的 495 个交易日中有 380 个样本通过显著性检验，即 $R^2 > R_{crit}^2$。其余 115 个，约 23.23%的样本缺少显著性。

我们对缺少显著性的样本进行仔细的观测，发现这些样本有两个特征：第一，相当一部分未通过显著性检验的成交量价分布，其成交价格太少，单日的样本数不足。这是因为我们在前期数据处理时只保留价格数据中小数点后的两位数，导致一些分布信息无法体现出来。为此，我们在小数点后第三位采用“二舍三入和七舍八入”的原则增加一个数值 0.005，相应的成交量也叠加到对应的交易价格中去，得到较明显的特征分布，并对它们进行拟合和显著性检验。结果有 28 天的分布样本通过零阶贝塞尔绝对值分布回归模型的检验。这样，我们总共有 408（380+28）个分布样本，约占总数的 82.42%，通过零阶贝塞尔绝对值分布回归模型的检验。

第二个特征是成交量在交易价格区间内出现两个或两个以上的峰值。这是因为这些样本在交易日当天的供求平衡关系发生较大的变化，其成交量价行为通过价格的调节，使得交易

价格从围绕某一定态均衡价格上下波动调整成围绕另外一个或几个定态均衡价格上下波动。定态均衡价格在当日出现不连续的跳跃变化。在这种情况下，其成交量价分布函数是（10）式的线性叠加，即

$$\left|\psi_m(p)\right| = \sum_n C_m \left|J_0\left[\omega_{m,n}(p - p_{0n})\right]\right| , \qquad (n = 1,2\cdots) \qquad （15）$$

其中 $n$ 是定态均衡价格的数目。我们采用具有两个定态均衡价格的量价分布回归模型

$$\left|\psi_m(p_i)\right| = \sum_i \sum_{n=1,2} C_m \left|J_0\left[\omega_{m,n}(p_i - p_{0,n})\right]\right| + \varepsilon_i \qquad (i = 1,2\cdots) \qquad （16）$$

对剩余的 87 个分布进行拟合和 $R_2^2 > R_{2crit}^2$ 显著性检验（这里 $k = 2$）。结果显示 59 个样本，约占总数的 11.92%在 95%的水平下显著成立（参见图 6（b））。

对还没有通过显著性检验的 28 分布样本，我们采用量价概率波分布的第二组解——含合流超几何函数（第一类 Kummer 函数）的量价分布回归模型来拟合。该函数的表达式如下

$$\left|\psi_m(p)\right| = C_m e^{-\sqrt{A_m}\left|p-p_0\right|} \cdot \left|F\left(-n,1,2\sqrt{A_m}\left|p - p_0\right|\right)\right| 。 \qquad （17）$$

由于这是一个多阶的分布函数，比较特殊和复杂，直接拟合比较困难，我们取定 $n = 1$ 时，采用一阶函数的展开式

$$\left|\psi_m(p)\right| = C_m e^{-\sqrt{A_m}\left|p-p_0\right|} \cdot \left|F\left(-1,1,2\sqrt{A_m}\left|p - p_0\right|\right)\right|$$

$$= C_m e^{-\sqrt{A_m}\left|p-p_0\right|} \cdot \left|1 - 2\sqrt{A_m}\left|p - p_0\right|\right| \qquad （18）$$

进行拟合和检验。结果显示 28 个样本中有 23 个（约占总分布数的 4.65%）在 95%的水平下显著成立（参见图 6（c））。最后 5 个分布样本呈现 4 个或 4 个以上的峰值，这表明当日市场非常不稳定（见图 6（d））。图 7 是分别为采用函数（10）、函数（16）和函数（18）拟合并检验其显著性的典型样本。

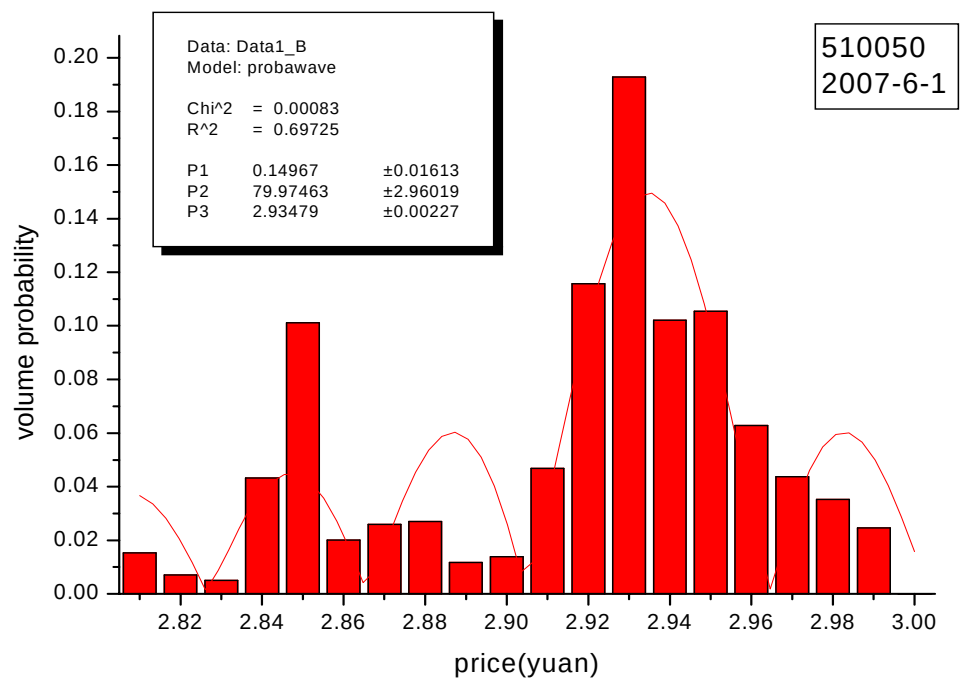

（a）采用函数（10）拟合的 2007-6-1 日数据

（$R^2 = 0.697 > R_{crit}^2 = 0.50$）

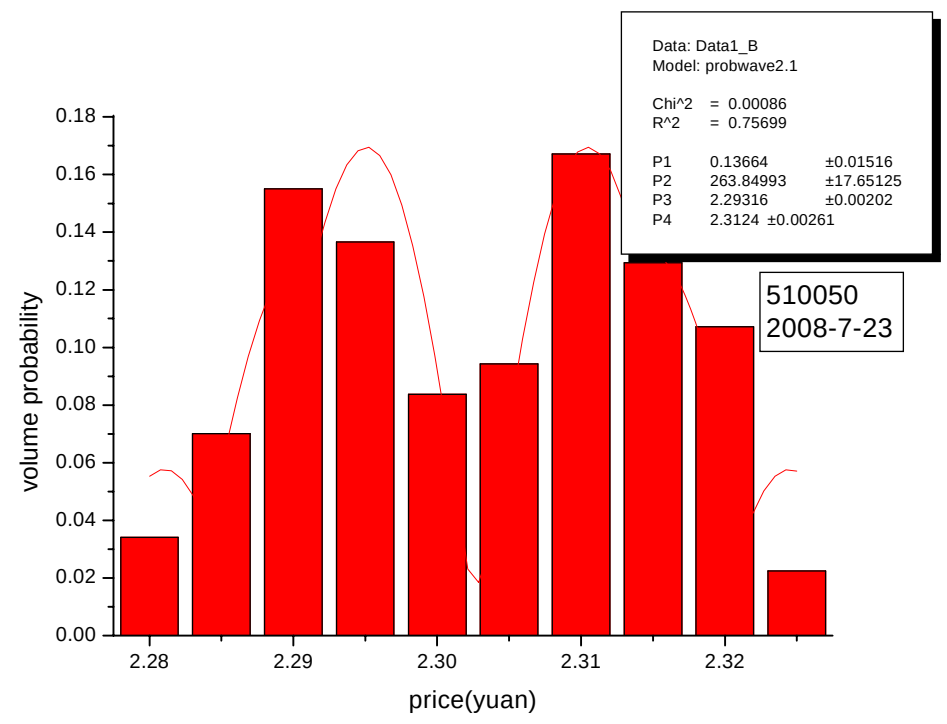

（b）采用函数（16）拟合的 2008-7-23 日数据

（$R_2^2 = 0.757 > R_{2crit}^2 = 0.575$）

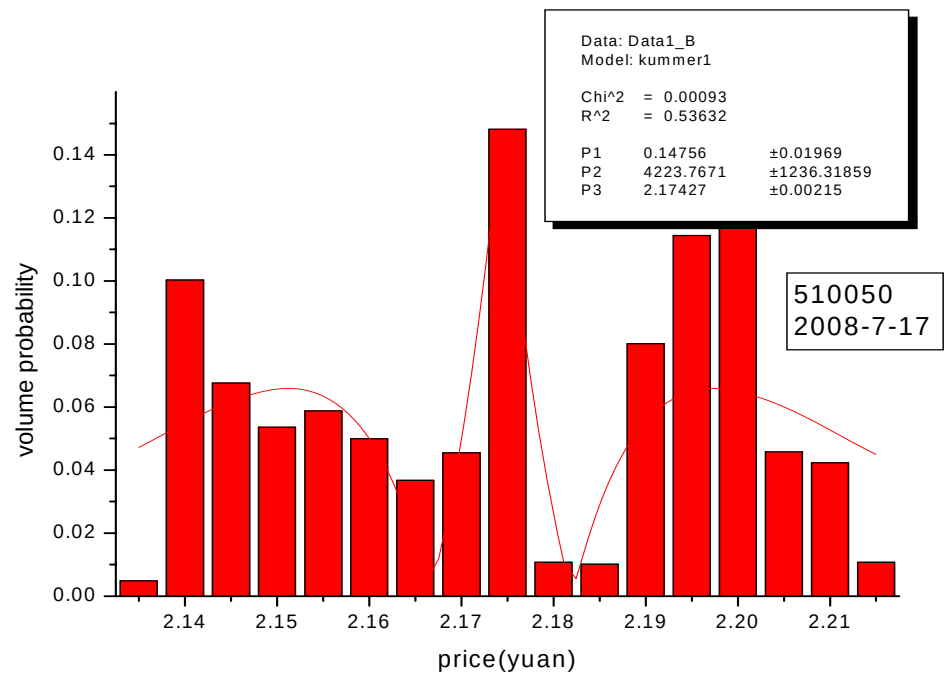

（c）采用函数（18）拟合的 2008-7-17 日数据

（$R^2 = 0.536 > R_{crit}^2 = 0.232$）

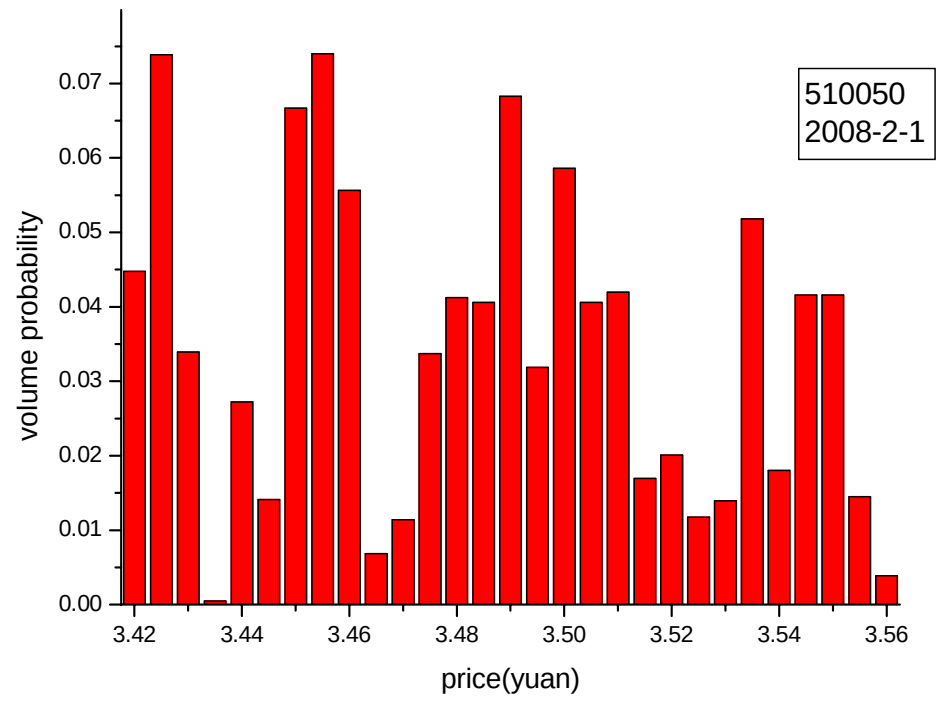

（d）5 个缺乏显著性的样本之一

图 6：用成交量价概率波回归模型拟合成交量在价格波动区间的分布并检验其显著性[1]

## 3.3 平均收益率与交易性条件反射强度变化率的相关性分析

对于 408 个通过零阶贝塞尔绝对值分布回归模型检验的样本（约占总数的 82.42%），我们从拟合数据中直接得到该交易日的定态均衡价格。对于其他 87 个交易日，我们用价格加权平均值近似表示该日的定态均衡价格。由此，我们能够计算出任意前后交易日之间定态均衡价格跳跃的幅度（平均收益率），研究平均收益率、交易性条件反射强度变化率以及成交金额变化率之间的相关性。在这里交易性条件反射强度变化率近似等于前后交易日总成交量变化率。

已知相关系数

$$r_{X,Y}=\frac{\operatorname{cov}(X,Y)}{\sigma_X\sigma_Y},\qquad（19）$$

其中$\sigma_X$和$\sigma_Y$分别是变量$X$和$Y$的标准差，$\operatorname{cov}(X,Y)$是协方差，$r_{X,Y}$是相关系数。我们采用$t$分布对相关系数进行显著性检验。建立原假设和备择假设

$$H_0:\rho=0\text{；}\ H_1:\rho\neq 0$$

计算统计量

$$t=\frac{|r-\rho|}{\sqrt{(1-r^2)/(n-2)}},\qquad（20）$$

其中$r$、$n$分别为样本相关系数和样本容量。取显著性水平为$\alpha=0.05$，当$t>t_{crit}=t_{0.05/2}(n-2)$时，拒绝原假设，相关系数在该显著性水平下显著不为 0。

我们使用 Eviews6.0 软件，以美国次贷金融危机和上证综合指数走势为背景，将华夏上证 50ETF（510050）整个样本数据从 2007 年 4 月 2 日至 2009 年 4 月 10 日又分成为五个子样本：2007 年 4 月 2 日（上证综合指数 3252.59 点）至 2007 年 6 月 29 日（上证综合指数 3820.70 点）为泡沫破裂前市场上升前期；2007 年 7 月 2 日（上证综合指数 3836.29 点）至 2007 年 10 月 31 日（上证综合指数 5954.77 点）为泡沫破裂前市场上升后期；2007 年 11 月 1 日（上证综合指数 5914.28 点）至 2008 年 4 月 30 日（上证综合指数 3693.11 点）为泡沫破裂后市场下跌前期；2008 年 5 月 5 日（上证综合指数 3761.01 点）至 2008 年 10 月 31 日（上证综合指数 1728.79 点）为泡沫破裂后市场下跌后期；2008 年 11 月 3 日（上证综合指数 1719.77 点）至 2009 年 4 月 10 日（上证综合指数 2444.23 点）为市场反转上升初期。

这样细分有一个优点：我们能够看出不同时期市场群体的心理行为随环境发生的变化。

从实证结果（参见表中的检验数据），我们看到：1）总体来说，平均收益率与交易性条件反射强度变化率和成交金额变化率之间都呈现出显著的正相关；2）在整个样本期间的 5 个不同阶段，平均收益率与交易性条件反射强度变化率之间的相关性明显不同:（a）在泡沫破裂前、后两个时期，它们之间的正相关缺乏显著性；(b) 在泡沫破裂后的后半时期，它们具有显著的正相关性；(c) 泡沫破裂后市场出现反转上升的初期，其正相关系数最高，是 0.4766；(d) 比较特殊的是在牛市上升的一段时期内，即在泡沫破裂前市场上升的前期，平均收益率与交易性条件反射强度变化率之间存在着显著的负相关，相关系数是-0.2567；3）成交金额变化率与平均收益率之间的相关系数比交易性条件反射强度变化率与平均收益率之间的相关系数总是要高出几个百分点。我们将在下一部分详细讨论实证结果。

[1] 在图 6 中，拟合结果中的P1、P2 和P3 分别代表归一化常数、本征值常数和定态均衡价格，在图 6（b）中，P3 和P4 分别代表跳跃前后的定态均衡价格。

表：相关系数和显著性检验

| | 时期 | 样本数量 | 上证综合指数（1A0001） | 收益率与条件反射变化率 | 收益率与成交额变化率 | 差值 | 条件反射变化率与成交额变化率 |
|---|---|---|---|---|---|---|---|
| A | 2007.4.2—2009.4.10 | 494 | 3252.59—2444.23 | 0.1391<br>（t=3.115> $t_{crit}$ =1.960） | 0.1970<br>（t=4.458> $t_{crit}$ =1.960） | 0.0579 | 0.9975<br>（t=316.3> $t_{crit}$ =1.960） |
| B | 2007.4.2—2007.6.29 | 59 | 3252.59—3820.70 | -0.2567<br>（t=2.006> $t_{crit}$ =2.001） | **-0.2120**<br>**（t=1.638< $t_{crit}$ =2.001）** | 0.0447 | 0.9986<br>（t=142.0> $t_{crit}$ =2.001） |
| C | 2007.7.2—2007.10.30 | 83 | 3836.29—5954.77 | **0.0729**<br>**（t=0.6583< $t_{crit}$ =1.990）** | **0.1053**<br>**（t=0.9529< $t_{crit}$ =1.990）** | 0.0324 | 0.9993<br>（t=241.2 > $t_{crit}$ =1.990） |
| D | 2007.11.1—2008.4.30 | 122 | 5914.28—3693.11 | **0.1026**<br>**（t=1.130< $t_{crit}$ =1.980）** | **0.1714**<br>**（t=1.906< $t_{crit}$ =1.980）** | 0.0688 | 0.9968<br>（t=137.0> $t_{crit}$ =1.980） |
| E | 2008.5.5—2008.10.31 | 123 | 3761.01—1728.79 | 0.1963<br>（t=2.202> $t_{crit}$ =1.980） | 0.2706<br>（t=3.091> $t_{crit}$ =1.980） | 0.0743 | 0.9958<br>（t=119.2> $t_{crit}$ =1.980） |
| F | 2008.11.3—2009.4.10 | 107 | 1719.77—2444.23 | 0.4766<br>（t=5.556> $t_{crit}$ =1.983） | 0.5203<br>（t=6.243> $t_{crit}$ =1.983） | 0.0437 | 0.9981<br>(t=166.5 > $t_{crit}$ =1.983) |

标注： 1） $t_{crit}$ 指 $t_{0.05/2}(n-2)$， $t > t_{crit}$ 说明相关系数显著不为零；相反，不能拒绝相关系数为零的原假设；

2） 差值是平均收益率与成交金额变化率之间的相关系数减平均收益率与交易性条件反射强度变化率之间的相关系数；

2） 红色黑体标出的是检验结果的相关性缺乏显著性；

3） 上证综合指数以当日收盘值计量。

## 四、实证分析和讨论

### 4.1 定态均衡和定态均衡价格

Shi（2006）将价格波动的行为分解成两个部分。第一，价格始终围绕某个定态均衡价格上下波动，这种定态行为可以用解析的成交量价概率波方程（5）定量描述，其中定态均衡价格是成交量峰值对应的交易价格；第二，定态均衡价格表现出不连续的跳跃（jump）变化。在定态均衡条件下，定态均衡回归力较弱，定态均衡容易被打破。当供需关系发生较大变化，例如有增量扰动资金进入市场大量买入股票，打破原有定态均衡，那么，供需关系就会通过价格的调节作用寻找新的定态均衡点，定态均衡价格出现不连续的跳跃。此后价格又开始围绕新的定态均衡价格上下波动。定态均衡价格跳跃的幅度表示价格波动的平均收益率。

我们对华夏上证 50ETF 近 740 天，共 495 个交易日，即 495 个成交量价分布分析，用零阶贝塞尔绝对值分布函数对它们逐一拟合和显著性检验，结果是通过显著性检验的样本数量是 408 个，占总数的 82.42%，这表明在股票市场中普遍存在着定态均衡状态。对于通过显著性检验的分布，我们从拟合数据中就能够得到交易日当天的定态均衡价格。没有通过显著性检验的成交量价分布“异常”现象是在交易日当天定态均衡价格跳跃的行为特征。对于这部分交易日，我们用成交量加权平均值近似代替其定态均衡价格。这样，我们能够得到前后交易日定态均衡价格跳跃的幅度和价格波动的平均收益率。

我们的实证进一步佐证了 Shi（2006）前期发现：在股票市场中普遍存在着定态均衡状态，而定态均衡价格会出现不连续的跳跃变化。

我们从本次“异常”样本数量比例（17.58%）与 Shi（2006）前期的比例（5.66%）比较，就能够推断 2003 年的上海证券交易市场比 2007~2009 年的市场平稳。事实上，上证指数在 2003 年最大振幅是 26.17%，而在 2007、2008 及 2009 年的最大振幅分别是 140.96%、231.71%和 88.60%。因此，我们能够用这种比值表示市场稳定性或波动性指标。比值越大，市场的波动性也就越大，越不稳定。

### 4.2 检验处置效应和羊群行为的新方法

前一部分的实证结果表明：从 2007 年 4 月到 2009 年 4 月整个样本期间，平均收益率与交易性条件反射变化率之间的相关系数是 0.1391，具有显著的正相关性（参见表中 A 栏）。为了理解它所包含的行为内容，我们先了解一下股市中的处置效应和羊群行为。

Shefrin 和 Statman（1985）最早提出处置效应：投资者在处置处于不同赢利状态股票时，存在急于兑现收益、回避兑现亏损的倾向。之后，Odean（1998）在研究了 10000 个帐户的交易记录后充分证实了处置效应，Weber 和 Camerer（1998）通过心理学实验来检验并且用 Kahneman 和 Tversky（1979，1992）的前景理论来解释处置效应，Grinblatt 和 Keloharju（2001）用芬兰股市中 293034 卖单数据检验了处置效应。处置效应表现的是卖股票时的行为异象。

羊群行为普遍存在于人类的社会活动中（Banerjee，1992），理论上讲与许多经济活动有关系（Graham，1999）。当模仿他人导致许多人采取同样的行为时，我们说羊群行为出现了。在股票市场中存在着许多类型的羊群行为（Hirshleifer 和 Teoh，2001），例如机构和个人的羊群行为（Nofsinger 和 Sias，1999）。Lux (1995)曾提出传染模型描述投资者的羊群行为。我们定义交易性条件反射的羊群行为是受价格波动和收益信息影响，预期收益的从众和规避亏损的观望交易行为。羊群行为表现的是买股票时的行为异象。

对于卖方，如果平均收益率正值越大卖量越多，并且如果平均收益率负值越大（亏损越大）卖量越少，那么，这是处置效应。对于买方，如果平均收益率正值越大，买量越多，从众跟风的行为越明显，并且如果平均收益率负值越大（亏损越多），买量越少，大家观望回避亏损的气氛越浓，这是交易性条件反射羊群行为。

因此，平均收益率与交易性条件反射强度变化率之间显著的正相关特性表明市场同时具有显著的处置效应和交易性条件反射羊群行为，其正相关系数的大小也能够用来衡量处置效应和羊群行为的强弱。我们的实证表明市场总体平均收益率与交易性条件反射变化率之间具有显

著的正相关性，具有显著的处置效应和羊群行为。它们是人类受价格波动和收益信息刺激、为获得收益和规避亏损从事交易而表现出来的一些生理反应和人性被收益条件化的行为。

### 4.3 认知心理和交易行为的变化

我们以美国次贷金融危机和我国股市泡沫破裂为背景，将整个样本时期细分 5 个阶段。这样做有一个优点，我们能够研究市场群体随价格波动和环境不同展现出的认知心理和交易行为的变化。

在泡沫破裂前的最后疯狂上涨和泡沫破裂初期，其相关系数分别是 0.0729 和 0.1026。它们的正相关都缺乏显著性，处置效应和羊群行为不显著。在这两个阶段，有限理性和非理性之间对价格认识的分歧增大，他们之间的博弈产生了很大的不确定性（参见表中 C 和 D 栏）。如果我们今后的研究能够进一步证明这种缺乏显著性相关特征与股市价格泡沫破裂有一定的相互联系，那么，我们就在一定程度上找到了一种预测和把握泡沫破裂的时机！

有趣的是在泡沫破裂后的后期，持续下跌和超跌期间，相关系数是 0.1963，具有显著的正相关性，处置效应和羊群行为显著增大，换句话说下跌越多，亏损越大，卖出越少，处置效应越显著；同时下跌越多，买入越少，观望气氛越浓，羊群行为更显著（比较表中 D 栏与 E 栏的相关系数）。

在反转上涨初期，相关系数最大，是 0.4766，具有显著的正相关性，处置效应和羊群行为最显著（参见表中 F 栏）。这说明两个问题：第一，经历一年大幅下跌，股指从最高的 6124.04 点跌到 1664.93 点后，人们已经形成了熊市下跌的趋势思维，短期获利了结的处置效应非常显著；第二，当下跌趋势形成之后，提高人们收益预期，保持持续增量资金入市接盘、克服大量短期赢利兑现的处置效应是改变市场继续下跌并出现反转的必要条件。

总体来说，在股票交易市场中普遍存在着处置效应和羊群行为，而且我们发现下跌越多，亏损越大，平均收益率与交易性条件反射强度变化率之间的正相关系数就越大，处置效应和羊群行为就越显著。

现在，我们来关注实证中非常特殊的情况，泡沫破裂前市场上升前期收益率与交易性条件反射强度的变化率之间存在较强的负相关性，相关系数是-0.2567，具有显著的负相关性（参见表中 B 栏）。

上证综合指数从 2005 年 6 月最低的 998.23 点开始几乎是一路上涨到 2007 年 3 月 30 日（样本选取区间起始日期）的 3183.98 点。当持有股票与正收益率挣钱（条件反射刺激物）不断结合，持有股票的行为转化为正收益率挣钱效应（正强化）时，一种新的条件反射也就形成了。人们表现出惜售和持有股票的心态。价格上升，卖的数量减少，成交量萎缩，而价格下跌，买的数量增加，成交量放大，出现“反处置效应”。

最后，我们还注意到平均收益率与成交金额变化率之间的相关性比它与交易性条件反射强度变化率之间的相关性总是高出几个点（0.03~0.08）。由此，我们可以推断在股票交易市场，资金对平均收益率的作用大于人们的心理行为和收益预期的作用，印证了华尔街的一句至理名言：现金为王。

### 4.4 交易性条件反射循环和过度交易

我们将传统的操作性条件反射的三项相互联系、可能发生的事件进行扩展和补充，归纳出交易性条件反射的六个环节：价格波动和收益信息（可识别刺激）→认知、判断和决策（决策）→交易或不交易（操作行为）→预期收益或亏损（结果）→损益结果影响情绪（新的可识别刺激）→调整收益预期（反馈）。我们将这种周而复始的过程称之为交易性条件反射行为的六项重复过程（参见图 4 和图 7）。

现在让我们回过头来再仔细思考一下引言中提到的颅内刺激（intra-cranial stimulation, ICS）实验。如果我们做一个类比，大鼠为获取食物而进行的操作性条件反射的行为是理性的，为获取愉快感而进行的操作性条件反射的过度行为是非理性的，那么，我们就能够类推人们过度交易而忽视了资产增值或效用最大化的非理性行为的原因。

在现代商品经济中，钱、收入和收益可用于交换商品和服务，也许是最重要的经济强化

物（Pierce 和 Cheney, 2004）。在股票市场中，交易性条件反射的强化物是钱和收益率，它不会像一级强化物那样因生理需求已经得到满足而失去强化作用；第二，量价概率波行为没有确定的时间周期，其收益率的不确定性增加了交易性条件反射的频率。收益对应与正强化物，亏损对应与负强化物。预期股票价格上涨会促使人们买入持有，预期股票价格下跌会促使人们卖出观望。因此，无论是预期价格上涨还是下跌都会促使人们交易，提高交易的频率，导致过度交易。第三，类似与操作性条件反射中部分强化的不定时间间隔模式（库恩，2007），持有股票时间有长有短之后才出现收益，这使得交易性条件反射消退的抑制力很强，例如投资者止损出来后很容易因为看到价格上涨导致收益预期变化再次买入股票（羊群行为），增加交易频率；第四，收益（盈利和亏损）结果会影响人们的情绪（emotion）和判断能力，并且再反馈到交易者、改变人们的收益预期。例如，投资者赢利卖出后会因为自信增强而再次参与买进股票，而刚买入股票之后也会因为收益预期变化或已经获得微利（处置效应）又急于卖掉，交易很难停止，出现过度交易。我们的结论是：交易性条件反射导致过度交易！

## 4.5 多学科交叉的金融理论体系

Shiller（2006）把过去半个世纪的金融理论史概括成两次革命。第一次起源于上个世纪六、七十年代的新古典金融学革命和第二次起源于上个世纪八十年代的行为金融学革命。与新古典金融学理论要求投资者是完全理性和决策效用最大化不同，行为金融学把新古典经济金融学理论与心理学和决策科学联系起来，研究投资者是如何产生系统的认知偏差、判断偏差和决策偏差，并从心理学的角度解释金融市场上的异常现象（Tversky 和 Kahneman，1973）。目前，行为金融学分别在研究人们偏好、情绪和信念、以及有限套利三方面对金融市场的影响已经取得了许多卓有成效的成果，其中由 Kahneman 和 Tversky（1979，1992）提出并且发展起来的描述性前景理论影响最深远。

然而，新兴发展中的行为金融学需要将心理学同经济学更好地结合起来构建比较统一的假设前提和研究范式，需要找到统一的理论将理性与非理性这两个极端的情况统一起来，以增强其解释能力。而目前的研究还局限在基于投资者的认知偏差和前景理论，但在各自的目标函数上引入了五花八门的行为特征，用以解释某一特定现象（董志勇，2009）。

Barberis 和 Thaler（2003）认为行为金融学模型有代表性地扑捉到投资信念（beliefs）、偏好（preferences）或有限套利（the limits of arbitrage）三个方面中的某一特征，而不能涵盖全部三个方面，而且对同一个实证结果又有明显不同的行为解释。

Shiller（2006）认为行为金融学不是完全不同与新古典金融学。也许描述两者不同的最好方式是行为金融学更加兼收并蓄、更加愿意从其他社会科学吸取营养而较少关心模型是否完美、更加注重描述真实的人类行为现象。

我们试图通过经济物理学的方法，将生理学、认知心理学、神经经济学（Camerer 等，2005）、医学、概率统计数学、经济金融学、行为金融学和经济物理学联系起来，在一个整体框架内系统全面地研究涉及多交叉学科的金融交易市场的行为和变化（参见图 7）。

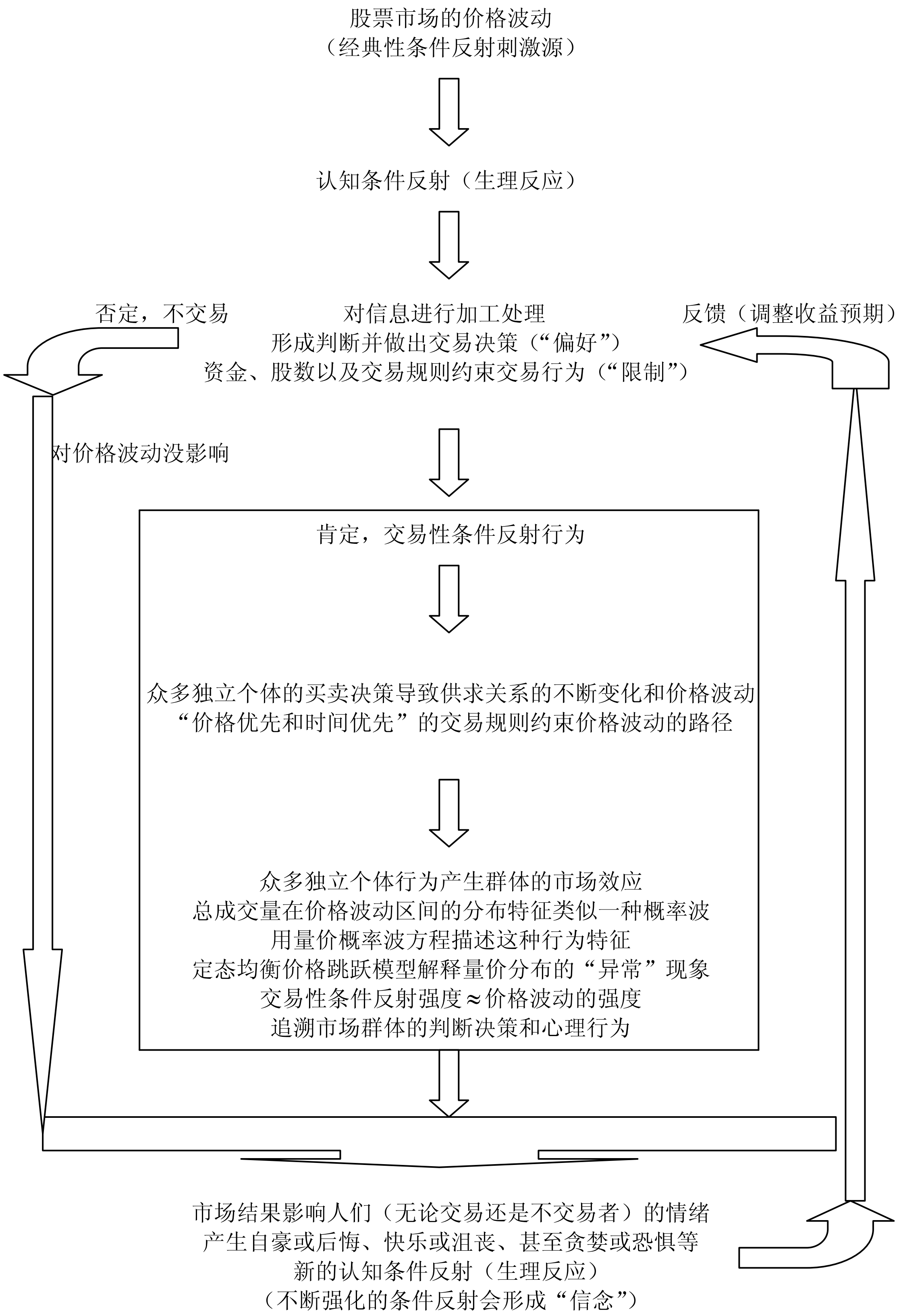

图 7：用成交量概率定量描述价格波动的不确定性和市场群体交易性条件反射的强度
（交易性条件反射导致过度交易）

### 4.6 应用前景

通过量价概率波分布函数，我们根据市场的心理行为用成交量概率定量描述了交易性条件反射的强度并且归纳出包含六项环节的交易性条件反射循环，用它解释过度交易行为，这有助于今后从认知心理学、神经经济学（Hsu 等，2005）和医学角度解释为什么不同性格的人会有不同的交易频率（交易量），并且检验过于自信、寻找兴奋感或意见分歧导致过度交易的行为金融学假说，例如，当价格上涨成交量增加的市场效应在一定程度上支持意见分歧导致过度交易假说（Hong 和 Stein，2007）；第二，通过对平均收益率与交易性条件反射强度变化率的相关性分析，我们用高频数据同时检验了市场群体的处置效应和羊群行为。交易性条件反射理论模型为行为金融学假说和模型提供了一个新的定量分析和检验方法；第三，如果我们今后的研究能够进一步证明平均收益率与交易性条件反射强度变化率之间相关性缺乏显著性的特征与股市价格泡沫破裂有一定的相互联系，那么，我们就在一定程度上找到预测和把握泡沫破裂的时机！通过加深对市场的认识，我们能够优化投资策略，规避和防范大的金融风险和扑捉好的投资机会；第四，有助于监管部门通过正负强化（reinforcement）措施和政策调控金融市场（Xiao 和 Houser，2005），降低大的金融风险和危机爆发的可能性；第五，证券市场为我们提供了一个很好的交易性条件反射实验室，有助于今后开展一些生理学、认知心理学、神经经济学和医学方面的研究工作。

## 五、总结和主要结论

在真实的金融交易市场中，有许许多多的因素都会影响人们的主观认知和成交量的变化。如何将市场群体的主观交易行为和成交量变化纳入到经济物理学中规范的解析量价概率波分布函数中，将是定量解决成交量、价格波动的平均收益率与增量资金之间相互关系的关键，也是建立关于价格波动和收益信息的交易性条件反射理论模型的关键。

我们从经典条件反射和操作性条件反射出发，进一步提炼交易性条件反射行为的定义：在股票市场中，交易性条件反射是一种受价格波动和收益信息影响，通过分析、处理、判断和决策进行交易，并预期未来收益的一种操作性条件反射；交易性条件反射结果反馈给交易者会影响他们的情绪和判断能力，调整他们的收益预期并进行交易活动。

基于解析的量价概率波分布函数——用成交量概率描述价格波动的不确定性和强度、用定态均衡价格跳跃的幅度描述价格波动的平均收益、并且用成交金额描价格波动和成交量变化的约束量，我们用成交量的变化率来计量交易性条件反射强度的变化率，构造了一个关于价格波动和收益信息的交易性条件反射的理论模型。

同理，当我们用定态均衡价格跳跃的幅度率表示前后交易日价格波动的平均收益率时，我们也能够用这两天成交量的变化率来描述市场群体对该平均收益率的交易性条件反射强度的变化率。这样，我们能够用高频数据，通过研究平均收益率与交易性条件反射强度变化率之间的相关性，追溯和分析市场群体在做交易决策时受价格波动和市场环境不同发生的心理行为的变化，试图为行为金融提供一个量化分析的研究方法和检验模型。

通过采用 2007 年 4 月到 2009 年 4 月我国股市中的华夏上证 50ETF 每笔交易的高频数据实证检验，我们首先来研究股票市场中的定态均衡，进一步佐证了 Shi（2006）前期的发现——在股票交易市场中，定态均衡状态普遍存在，量价概率波分布函数是有效的；第二，通过高频数据研究平均收益率与交易性条件反射强度变化率之间的相关性，我们得到许多有意义的结果。这些包括：1）我们用它们之间的正相关性同时检验和解释了市场群体的处置效应和羊群行为；总体来说，市场同时具有显著性的处置效应和羊群行为。我们用交易性条件反射理论来解释这些市场群体行为的异象：它们是人类受价格波动和收益信息刺激、为获得收益和规避亏损从事买卖交易而表现出来的一些生理反应和人性被收益条件化的行为；2）发现了市场中的反处置效应，并且用牛市中买入持有行为不断获得收益的正强化条件反射解释这种市场异象；3）当下跌趋势形成之后，提高人们收益预期，保持持续增量资金入市接盘、克服大量短期赢利兑现的处置效应是改变市场继续下跌并出现反转的必要条件；4）在泡沫破裂前后的两个时期，它们的正相关都缺乏显著性，这也许可以帮助我们把握泡沫破裂

的时机；5）成交金额变化率与收益率之间的相关系数比交易性条件反射强度变化率与它之间的相关系数总是要高出几个百分点。我们由此能够推断：在股票交易市场，资金对收益率的作用大于人们的心理行为和损益预期的作用；第三，交易性条件反射导致过度交易。

参考文献：

Ait-Sahalia, Yacine (2004): “Disentangling Diffusion from Jump,” *Journal of Financial Economics*, 74, 487-528.

Banerjee, Abhijit V. (1992): “A Simple Model of Herd Behavior,” *Quarterly Journal of Economics*, CVII, Issue 3, 797-817.

Benos, Alexandros V. (1998): “Aggressiveness and Survival of Overconfident Traders,” *Journal of Financial Markets*, 1, 353-383

Barber, Brad M., AND Terrance Odean (2000): “Trading Is Hazardous to Your Wealth: The Common Stock Investment Performance,” *Journal of Finance*, 55, 773-806.

Barber, Brad M., Terrance Odean, AND Ning Zhu (2009): “Systematic Noise,” *Journal of Financial Markets*, 12, 547-569.

Barberis, Nicholas, AND Richard Thaler (2003): “A Survey of Behavioral Finance,” *Handbook of the Economics of Finance (Edited by G.M. Constantinides, M. Harris and R. Stulz),* Elsevier Science B.V., 1051-1121.

Camerer, Colin F., George Loewenstein, AND Drazen Prelec (2005): “Neurosciences: How neuroscience can inform economics?” *Journal of Economic Literature*, XLIII, 9-64.

Cowles, J.T. (1937): “Food-tokens as Incentive for Learning by Chimpanzees,” *Comparative Psychology Monographs*, 14 (5, Whole no. 71).

Dragoi, V. (1997): “A Dynamic Theory of Acquisition and Extinction in Operant Learning,” Neural Networks, 10, 201-229.

Graham, John R. (1999): “Herding among Investment Newsletters: Theory and Evidence,” *Journal of Finance*, 54, 237-268.

Graham, John R., Campbell R. Harvey, AND Hai Huang (2009): “Investor Competence, Trading Frequency, and Home Bias,” *Management Science*, 55, 1094-1106.

Grinblatt, Mark, AND Matti Keloharju (2001): “What Makes Investors Trade?” *Journal of Finance*, 56, 589-616.

Grinblatt, Mark, AND Matti Keloharju (2009): “Sensation Seeking, Overconfidence, and Trading Activity,” *Journal of Finance*, 64, 549-578.

Hirshleifer, David AND Siew Hong Teoh (2009): “Thought and Behavior Contagion in Capital Markets,” Handbook of Financial Markets: Dynamics and Evolution (edited by Hens and Schenk-Hoppe), North-Holland/Elsevier.

Hong, Harrison, AND Jeremy C. Stein (2007): “Disagreement and Stock Market,” *Journal of Economic Perspectives*, 21, 109-128.

Hsu, Ming, Meghana Bhatt, Ralph Adolphs, Daniel Tranel, AND Colin F. Camerer (2005): “Neural Systems Responding to Degrees of Uncertainty in Human Decision-Making,” *Science*, 310 (5754), 1680-1683.

Irons, Jessica G., and William Buskist (2008): “Operant Conditioning,” 21st Century Psychology—A Reference Handbook (ed. by Davis and Buskist), 329-339.

Kahneman, Daniel, AND Amos Tversky (1979): “Prospect Theory: An Analysis of Decision under Risk,” *Econometrica*, 47, 263-292.

Lee, Charles M.C., AND Bhaskaran Swaminathan (2000): “Price Momentum and Trading Volume,” *Journal of Finance*, 55, 2017-2069.

Li, Haitao, Martin T. Wells, AND Cindy L. Yu (2008): “A Bayesian Analysis of Return Dynamics with Levy Jumps,” *The Review of Financial Studies*, 21, No.5, 2345-2378.

Lo, Andrew W., AND Jiang Wang (2006): “Trading volume: Implications of an Intertemporal Capital Asset Pricing Model,” *Journal of Finance*, 61, 2805-2840.

Lux, Thomas (1995): “Herd Behavior, Bubbles and Crashes,” *The Economic Journal*, 105 (July), 881-896.

McCauley J.L. (2000): “The Futility of Utility: How Market Dynamics marginalize Adam Smith,” *Physica A*, 285, 506-538.

Nofsinger, John R. AND Richard W. Sias (1999): “Herding and Feedback Trading by Institutional and Individual Investors,” *Journal of Finance*, 54 (6), 2263-2295.

Odean, Terrance (1998a): “Volume, Volatility, Price, and Profit When All Traders Are Above Average,” *Journal of Finance*, 53, 1887-1934.

Odean, Terrance (1998b): “Are Investors Reluctant to Realize Their Losses?” *Journal of Finance*, 53, 1775-1798.

Odean, Terrance (1999): “Do Investors Trade Too Much?” The American Economic Review, 89, 1279-1298.

Olds, M.E. AND J.L. Fobes (1981): “The Central Basis of Motivation: Intracranial Self-stimulation Studies,” *Annual Review of Psychology*, 32, 523-574.

Osborne, M. F. M. (1977): The Stock Market and Finance from a Physicist’s Viewpoint, Grossga, Mineapolis.

Pavlov, Ivan (1904): “Physiology of Digestion,” *Nobel Lectures—Physiology or Medicine 1901-1921*, Elsevier Publishing Company, Amsterdam, 1967.
(available at http://nobelprize.org/nobel_prizes/medicine/laureates/1904/pavlov-lecture.html )

Pierce, W. D., AND C. D. Cheney (2004): Behavior Analysis and Learning (3rd.), Mahwah, NJ: Erlbaum.

Sewell, Martin (2008): “Behavioral Finance,” working paper.

Shefrin, Hersh, AND Meir Statman (1985): “The Disposition to Sell Winners Too Early and Ride Losers to Long: Theory and Evidence,” *Journal of Finance*, 40, 777-790.

Shi, Leilei (2006): “Does Security Transaction Volume-price Behavior resemble a Probability Wave?” *Physica A*, 366, 419-436.

Shiller，Robert J. (2006): “Tools for Financial Innovation: Neoclassical versus Behavioral Finance,” *The Financial Review*, 41, 1-8.
Skinner, B.F. (1938): The Behavior of Organisms: An Experimental Analysis, New York: Appleton-Century-Crofts
Statman, Meir, Steven Thorley, AND Keith Vorkink (2006): “Investor Overconfidence and Trading Volume,” *The Review of Financial Studies*, 19, 1531-1565.
Thorndike, E.L. (1913): Educational Psychology (Vol. 2), New York: Teachers College.
Tversky, Amos, AND Daniel Kahneman (1973): “Availability: A Heuristic for Judging Frequency and Probability,” *Cognitive Psychology*, 5, 207-232.
Tversky, Amos, AND Daniel Kahneman (1992): “Advances in Prospect Theory: Cumulative Representation of Uncertainty,” Journal of Risk and Uncertainty,” 5, 297-323.
Weber, Martin, AND Colin F. Camerer (1998): “The disposition effect in securities trading: an experimental analysis,” *Journal of Economic Behavior and Organization*, 33, Issue 2, 167-184.
Xiao, Erte, AND Daniel Houser (2005): “Emotion Expression in Human Punishment Behavior,” Proceedings of the National Academy of Sciences of the United States of America, 102 (20), 7398-7401.
董志勇（2009）:《行为金融学》，北京大学出版社，303-304.
库恩（Coon 美）等著，郑钢等译 （2007）:《心理学导论——思想与行为的认识之路》，第 11 版，中国轻工业出版社，283-323.
索罗斯（Soros 美）著，余济群、黄嘉斌译（1998）:《金融炼金术》，吉林人民出版社，3-57.
周衍柏（1985）:《理论力学教程》(第二版)，高等教育出版社，267-272.